\documentclass[useAMS,usenatbib]{mn2e}
\usepackage{epsf,graphicx,rotating,url}

\title[Near-IR BELR of AGN -- II. The 1~$\mu$m continuum]{The near-infrared broad emission line region of active galactic nuclei -- II. The one-micron continuum}

\author[H. Landt et al.]{Hermine Landt$^1$\thanks{E-mail: hlandt@unimelb.edu.au}\thanks{Visiting Astronomer at the Infrared Telescope Facility, which is operated by the University of Hawaii under Cooperative Agreement no. NNX-08AE38A with the National Aeronautics and Space Administration, Science Mission Directorate, Planetary Astronomy Program.}, Martin Elvis$^2$\footnotemark[2], Martin J. Ward$^3$\footnotemark[2], Misty C. Bentz$^4$\footnotemark[2]\thanks{Current address: Department of Physics and Astronomy, Georgia State University, 709 One Park Place South, Atlanta, GA 30303, USA}, 
\newauthor
Kirk T. Korista$^5$ and Margarita Karovska$^2$ \\ 
$^1$School of Physics, University of Melbourne, Parkville, VIC 3010, Australia \\ 
$^2$Harvard-Smithsonian Center for Astrophysics, 60 Garden Street, 
Cambridge, MA 02138, USA \\
$^3$Department of Physics, University of Durham, South Road, Durham, DH1 3LE \\ 
$^4$Department of Physics and Astronomy, University of California Irvine, 
4129 Frederick Reines Hall, Irvine, CA 92697, USA \\
$^5$Department of Physics, Western Michigan University, 
1903 W. Michigan Avenue, Kalamazoo, MI 49008, USA}

\begin{document}

\def\la{\mathrel{\hbox{\rlap{\hbox{\lower4pt\hbox{$\sim$}}}\hbox{$<$}}}}
\def\ga{\mathrel{\hbox{\rlap{\hbox{\lower4pt\hbox{$\sim$}}}\hbox{$>$}}}}

\font\sevenrm=cmr7
\def\OIII{[O~{\sevenrm III}]}
\def\SIII{[S~{\sevenrm III}]}
\def\FeII{Fe~{\sevenrm II}}

\def\cloudy{{\sevenrm CLOUDY}}

\date{Accepted ~~. Received ~~; in original form ~~}

\pagerange{\pageref{firstpage}--\pageref{lastpage}} \pubyear{2011}

\maketitle

\label{firstpage}

\begin{abstract}

  We use quasi-simultaneous near-infrared (near-IR) and optical
  spectroscopy from four observing runs to study the continuum around
  1 $\mu$m in 23 well-known broad-emission line active galactic nuclei
  (AGN). We show that, after correcting the optical spectra for host
  galaxy light, the AGN continuum around this wavelength can be
  approximated by the sum of mainly two emission components, a hot
  dust blackbody and an accretion disc. The accretion disc spectrum
  appears to dominate the flux at $\sim 1$~$\mu$m, which allows us to
  derive a relation for estimating AGN black hole masses based on the
  near-IR virial product. This result also means that a near-IR
  reverberation programme can determine the AGN state independent of
  simultaneous optical spectroscopy. On average we derive hot dust
  blackbody temperatures of $\sim 1400$~K, a value close to the
  sublimation temperature of silicate dust grains, and relatively low
  hot dust covering factors of $\sim 7\%$. Our preliminary variability
  studies indicate that in most sources the hot dust emission responds
  to changes in the accretion disc flux with the expected time lag,
  however, a few sources show a behaviour that can be attributed to
  dust destruction.

\end{abstract}

\begin{keywords}
galaxies: active -- galaxies: nuclei -- infrared: galaxies -- quasars: general
\end{keywords}

\section{Introduction}

The broad emission line region (BELR) of active galactic nuclei (AGN)
is one of the most direct tracers of the immediate environment of
supermassive black holes. However, despite decades of intensive
optical and ultraviolet (UV) spectrophotometric studies its geometry
and kinematics remain ill-defined \citep[see, e.g., review
by][]{Sul00}. Our current, limited knowledge of its physical condition
and scale was gained primarily through the application of
photoionisation models \citep[see, e.g., review by][]{Ferl03} and
through reverberation mapping studies \citep[see, e.g., review
by][]{Pet93}. We have started to extend these studies to near-infrared
(near-IR) wavelengths. In \citet[][hereafter Paper I]{L08a} we
outlined the rationale of our programme, presented the observations of
the first three epochs and addressed briefly some of the important
issues regarding the physics of the most prominent broad emission
lines. Here we present the fourth epoch of observation and investigate
the continuum around the 1 $\mu$m inflection point.

The AGN spectral continuum region around the rest-frame wavelength of
$\sim 1$ $\mu$m is believed to sample simultaneously two important
emission components, namely, the accretion disc \citep[e.g.,][]{Mal82,
  Mal83} and the hottest part of the putative dusty torus
\citep[e.g.,][]{Barv87, Neu87}. However, although it is assumed to be
understood, it has not yet been sampled spectroscopically in its
entirety. By probing the long-wavelength end of the accretion disc
spectrum \citep{Kish05, Kish08} such an investigation has the
potential to solve the discrepancy often found between theoretical
models and observations \citep[see, e.g., review
by][]{Korat99}. Furthermore, since the accretion disc most likely
illuminates directly the inner parts of the dust structure, monitoring
the change in spectral flux and slope of these two emission components
relative to each other can constrain the location and geometry of the
obscurer. So far, only few sources have been observed in dust
reverberation programmes \citep[e.g.,][]{Glass92, Nel96, Okn99,
  Glass04, Min04, Sug06, Kosh09}.

The paper is organised as follows. In Section 2 we briefly introduce
the sample and discuss the observations. In Section 3 we derive pure
AGN continuum spectral energy distributions (SEDs), based on which we
constrain the individual continuum components (Section 4). The
variability of these components is disussed in Section 5. Finally, in
Section 6 we summarize our main results and present our
conclusions. Throughout this paper we have assumed cosmological
parameters $H_0 = 70$ km s$^{-1}$ Mpc$^{-1}$, $\Omega_{\rm M}=0.3$,
and $\Omega_{\Lambda}=0.7$.

\section{The observations}

\begin{table*}
\caption{\label{irtflog} 
IRTF Journal of Observations for 4th Epoch}
\begin{tabular}{llrcrrrlrc}
\hline
Object Name & observation & exposure & airmass & \multicolumn{3}{c}{continuum S/N} 
& \multicolumn{3}{c}{telluric standard star} \\
& date & [sec] && J & H & K & name & distance & airmass \\
&&&&&&&& [deg] & \\
(1) & (2) & (3) & (4) & (5) & (6) & (7) & (8) & (9) & (10) \\
\hline
Mrk 335         & 2007 Jan 25 & 32x120 & 1.431 &  33 & 110 & 272 & HD1160   & 16.1 & 1.977 \\
Mrk 590         & 2007 Jan 24 & 58x120 & 1.214 &  37 & 122 & 143 & HD13936  & 10.2 & 1.697 \\
Ark 120         & 2007 Jan 26 & 48x120 & 1.162 & 187 & 226 & 436 & HD34317  &  1.8 & 1.054 \\
Mrk 79          & 2007 Jan 25 & 48x120 & 1.334 &  97 & 202 & 448 & HD45105  & 12.6 & 1.128 \\
PG 0844$+$349   & 2007 Jan 24 & 48x120 & 1.138 &  99 & 170 & 259 & HD71906  &  4.2 & 1.271 \\
Mrk 110         & 2007 Jan 26 & 48x120 & 1.449 &  61 & 148 & 452 & HD71906  & 17.8 & 1.368 \\
NGC 3227        & 2007 Jan 25 & 16x120 & 1.339 &  57 &  94 & 146 & HD89239  &  7.6 & 1.122 \\
NGC 4151        & 2007 Jan 24 &  8x120 & 1.446 &  15 &  54 & 113 & HD109615 &  4.9 & 1.398 \\
3C 273          & 2007 Jan 25 & 40x120 & 1.060 & 123 & 343 & 347 & HD109309 & 11.6 & 1.199 \\
HE 1228$+$013   & 2007 Jan 25 & 48x120 & 1.370 &  40 & 101 & 274 & HD109309 & 10.7 & 1.242 \\
NGC 4593        & 2007 Jan 24 &  8x120 & 1.446 &  43 & 123 & 191 & HD112304 & 10.8 & 1.540 \\
NGC 5548        & 2007 Jan 24 & 16x120 & 1.319 &  57 &  73 & 145 & HD131951 & 14.0 & 1.299 \\
Mrk 817         & 2007 Jan 26 & 40x120 & 1.509 &  98 & 224 & 323 & HD121409 &  7.8 & 1.480 \\
\hline
\end{tabular}

\parbox[]{14.3cm}{The columns are: (1) object name; (2) observation date; (3)
  exposure time; (4) average airmass; S/N in the continuum over $\sim
  100$~\AA~measured at the central wavelength of the (5) J, (6) H, and
  (7) K band; for the star used to correct for telluric absorption (8) name, 
  (9) distance from the source, and (10) average airmass.}

\end{table*}

\begin{table*}
\caption{\label{flwolog} 
Tillinghast Journal of Observations for 4th Epoch}
\begin{tabular}{lllccl}
\hline
Object Name & IRTF & observation & exposure & airmass & cloud \\ 
& observation & date & [sec] & & condition \\
& date &&&& \\
(1) & (2) & (3) & (4) & (5) & (6) \\
\hline
Mrk 335         & 2007 Jan 25 & 2007 Jan 24 & 2x300 & 1.38 & clear  \\
Mrk 590         & 2007 Jan 24 & 2007 Jan 24 & 2x720 & 1.21 & clear  \\
Ark 120         & 2007 Jan 26 & 2007 Feb 08 & 2x480 & 1.18 & cirrus \\
Mrk 79          & 2007 Jan 25 & 2007 Jan 22 & 2x480 & 1.12 & clear  \\
PG 0844$+$349   & 2007 Jan 24 & 2007 Jan 22 & 2x480 & 1.03 & clear  \\
Mrk 110         & 2007 Jan 26 & 2007 Jan 22 & 2x420 & 1.08 & clear  \\
NGC 3227        & 2007 Jan 25 & 2007 Jan 22 & 2x180 & 1.02 & clear  \\
NGC 4151        & 2007 Jan 24 & 2007 Feb 12 & 2x 90 & 1.07 & cloudy \\
3C273           & 2007 Jan 25 & 2007 Feb 12 & 2x300 & 1.25 & cloudy \\
HE 1228$+$013   & 2007 Jan 25 & 2007 Feb 12 & 2x480 & 1.30 & cloudy \\
NGC 4593        & 2007 Jan 24 & 2007 Feb 17 & 2x270 & 1.27 & cirrus \\
NGC 5548        & 2007 Jan 24 & 2007 Feb 17 & 2x270 & 1.02 & cirrus \\
Mrk 817         & 2007 Jan 26 & 2007 Feb 17 & 2x360 & 1.13 & cirrus \\
\hline
\end{tabular}

\parbox[]{10.7cm}{The columns are: (1) object name; (2) IRTF observation 
  date (reproduced from Table \ref{irtflog}); (3) Tillinghast UT
  observation date; (4) exposure time; (5) average airmass; and (6)
  cloud condition.}

\end{table*}

The target selection, observational strategy, and data reduction
procedures have been described in detail in Paper I. In short, we
obtained for a sample of 23 well-known relatively nearby ($z\la0.3$)
and bright ($J\la14$ mag) broad-emission line AGN during four
observing runs contemporaneous (within two months) near-IR and optical
spectroscopy. The observations were carried out between 2004 May and
2007 January with a single object being typically observed twice
within this period.

In the near-IR we used the SpeX spectrograph \citep{Ray03} at the NASA
Infrared Telescope Facility (IRTF), a 3 m telescope on Mauna Kea,
Hawai'i. We chose the short cross-dispersed mode (SXD, $0.8-2.4$
$\mu$m) and a short slit of $0.8\times15''$. The optical spectra were
obtained with the FAST spectrograph \citep{Fast98} at the Tillinghast
\mbox{1.5 m} telescope on Mt. Hopkins, Arizona, using the 300 l/mm
grating and a $3''$ long-slit, resulting in a wavelength coverage of
$\sim 3720-7515$~\AA.

Paper I presented details for the first three epochs (2004 May, 2006
January, and 2006 June). Similarly, we give in Tables \ref{irtflog}
and \ref{flwolog} the details for the fourth IRTF epoch carried out on
2007 January 24-26. These nights were mostly photometric with a seeing
in the range of $\sim 0.7 - 1''$.

\section{AGN spectral energy distributions} \label{agnsed}

\begin{figure}
\centerline{
\includegraphics[scale=0.4]{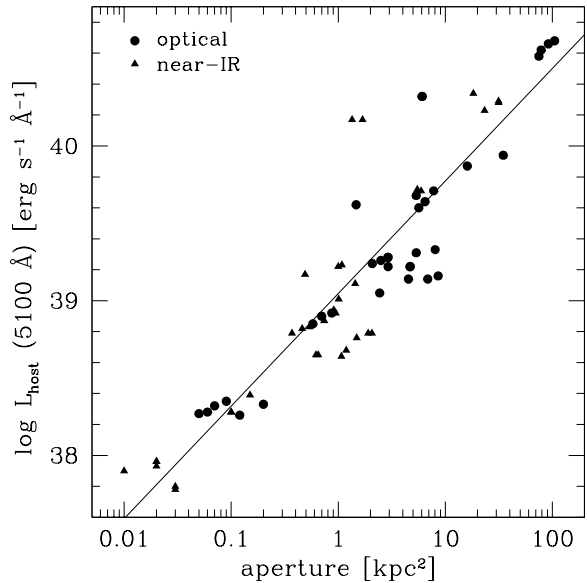}
}
\caption{\label{host} Spectral aperture versus enclosed host galaxy
  luminosity at rest-frame 5100~\AA~as derived from {\sl HST}
  images. Based on the observed correlation (solid line) we have
  estimated the host galaxy flux in the spectra of sources without
  useful {\sl HST} images.}
\end{figure}

\begin{figure}
\centerline{
\includegraphics[scale=0.4]{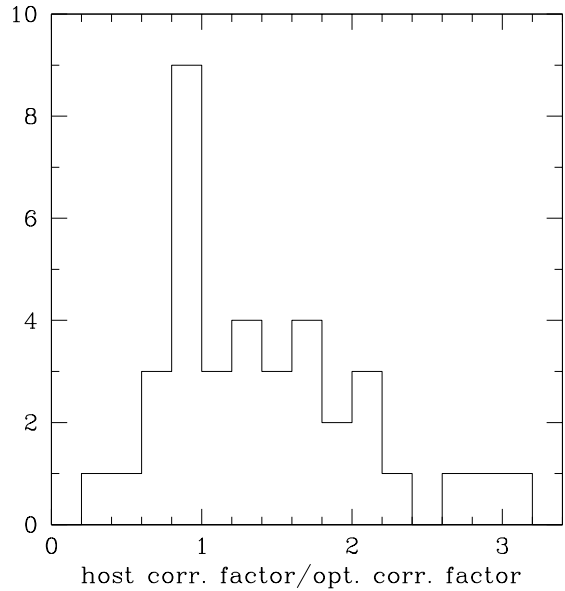}
}
\caption{\label{corfac} Histogram of the ratio between the optical
  correction factors applied to the host galaxy and the total flux.}
\end{figure}

\begin{figure}
\centerline{
\includegraphics[scale=0.4]{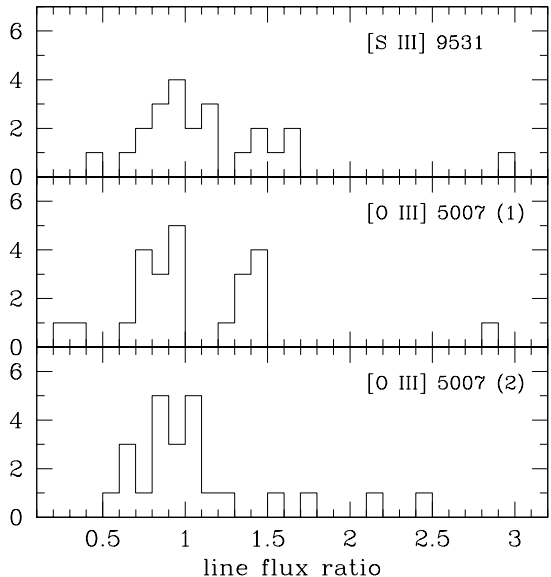}
}
\caption{\label{oiii} Histograms of the ratio between the line fluxes
  corresponding to the best-weather IRTF run and another epoch. The
  top, middle and bottom panels show the narrow emission lines
  \SIII~$\lambda 9531$, \OIII~$\lambda 5007$, and \OIII~$\lambda 5007$
  with the optical correction factor listed in Table \ref{hostopt}
  applied, respectively.}
\end{figure}

\begin{figure}
\centerline{
\includegraphics[scale=0.4]{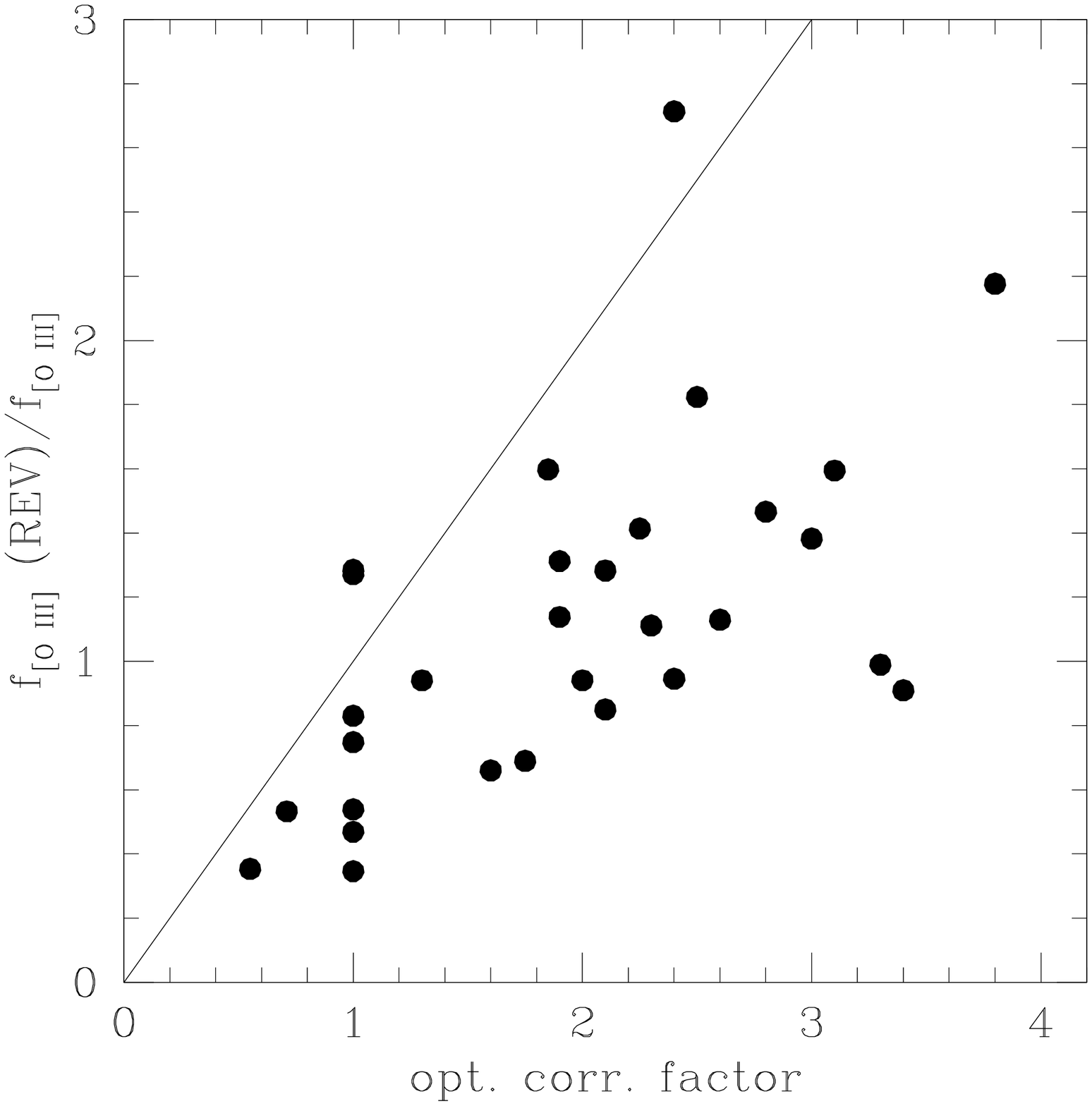}
}
\caption{\label{reverb} The ratio between the \OIII~$\lambda 5007$
  line fluxes from optical reverberation programmes and our observations
  versus the correction factor applied to the optical spectrum. The
  locus of equality is shown as the solid line.}
\end{figure}

Our aim is to study the spectral shape of the AGN continuum over the
entire observed frequency range and how it changes as the AGN flux
varies. For this purpose we need to isolate the spectrum of the pure
AGN and ensure an acceptable relative photometry between epochs. In
practice this means that we need to subtract the flux of the
underlying host galaxy and correct the individual spectra for flux
calibration errors (e.g., varying weather conditions, aperture and
seeing effects). In the absence of short-term variability, these steps
should yield an alignment between the AGN optical and near-IR spectral
parts.

In principle, accurate (within a few per cent) relative photometry of
the spectra can be achieved a posteriori by using the fact that the
narrow emission line fluxes of AGN are constant over several years
\citep{Pet88, Pet93}. However, this requires that the observations
were performed under similar conditions, i.e., using the same aperture
and slit position angle and having the same seeing. In general, these
requirements were not met for our optical observations, which were
obtained in service-mode under widely varying weather conditions. On
the other hand, the near-IR spectra were obtained uniformly during
four separate epochs, with one of them (2006 June) having photometric
weather conditions throughout. In addition, the much better seeing
achieveable in the near-IR relative to the optical band allowed us to
use a relatively small slit for these spectra, which minimized the
contamination by host galaxy light.

Given these considerations, we chose the following approach to obtain
pure AGN continuum SEDs with acceptable relative photometry. Firstly,
we performed relative photometry of the near-IR spectra using the flux
of the strong narrow emission line \SIII~$\lambda 9531$ with the epoch
observed under the best weather conditions (usually 2006 June or 2007
January) as a reference. We have measured the emission line fluxes as
described in Paper I and list them in Table \ref{hostir}. We note that
\SIII~$\lambda 9531$ is not observed for the source HE~1228$+$013, and
therefore, no adjustment of the near-IR spectra was made in this
case. Secondly, we subtracted the host galaxy flux from the (adjusted)
near-IR spectra. Finally, we corrected the flux scale of the optical
spectra by requiring that after the subtraction of the (optical) host
galaxy contribution the optical and near-IR AGN spectral parts
align. In Fig. \ref{SEDsub} we show the original data (left panels)
and the data after host galaxy subtraction and spectral alignment were
applied (right panels).

We have estimated the host galaxy contribution in the apertures of
both the near-IR and optical spectra using the {\sl Hubble Space
  Telescope (HST)} images of \citet{Bentz06a} and \citet{Bentz09} and
following their approach. The observed {\sl HST} fluxes were
transformed to a rest-frame wavelength of 5100~\AA~by applying a color
correction factor based on the model bulge galaxy template of
\citet{Kin96} and were corrected for Galactic extinction using
$A_{\lambda}$ values derived from the hydrogen column densities of
\citet{DL90}. The unabsorbed rest-frame 5100~\AA~fluxes were then used
to scale the galaxy template from \citet{Man01} of the appropriate
Hubble type, which was subsequently subtracted from the (rest-frame)
spectrum. Details of our host galaxy flux estimates are given in
Tables \ref{hostir} and \ref{hostopt} for the near-IR and optical
spectra, respectively. As expected, the relatively small aperture used
in the near-IR includes significantly less host galaxy flux than the
optical aperture, on average a factor of $\sim 3$.

For 8/23 objects we did not have suitable {\sl HST} images. In these
cases we have estimated the host galaxy contribution based on the
linear correlation present between the logarithms of the aperture and
the enclosed host galaxy luminosity for the sources with {\sl HST}
images (Fig. \ref{host}). The observed correlation is $\log L_{\rm
  host} = 0.73 (\pm 0.04) \cdot \log ({\rm aperture}) +
39.05(\pm0.04)$, where $L_{\rm host}$ is the host galaxy luminosity at
rest-frame 5100~\AA~(in erg s$^{-1}$ \AA$^{-1}$) and aperture is the
spectral aperture (in kpc$^2$). Additionally, we could not find
information on the host galaxy type for four sources. In these cases
we have assumed elliptical hosts for the two high-redshift sources
IRAS 1750$+$508 and PDS 456 and spiral (S0) hosts for the two
low-redshift sources H~2106$-$099 and H~1934$-$063.

The frequency gap between our near-IR and optical spectra is large
enough ($\log \nu \sim 0.05$) that we need to assume an overall
spectral shape in order to judge the two parts as being `aligned'.
Therefore, since the AGN continuum blueward of $\sim 1$ $\mu$m is
generally thought to be emitted by the accretion disc, we have
considered this component as the model and calculated its spectrum as
detailed in Section \ref{accretion}. Assuming an accretion disc
spectrum, excess host galaxy contribution to the optical spectrum (but
not to the near-IR spectrum) was apparent in the majority of our low
redshift ($z\la0.1$) sources, requiring also a host galaxy flux
correction factor. However, as Fig. \ref{corfac} shows, in most cases
the host correction factors are similar to the correction factors
applied to the total optical spectra, and in only a few sources they
indicate a real underestimation (by factors of $\la 3$).
 
The relative photometry in the near-IR required flux correction
factors of $<2$, with the exception of Mrk 590 (factor of $\sim 3$),
and most spectra had flux differences relative to the best-weather
epoch of no more than $\sim 20\%-30\%$ (Fig. \ref{oiii}, top panel).
The correction factors for the optical spectra spanned a larger range
of $\sim 0.6 - 3.8$. The narrow emission line \OIII~$\lambda 5007$ is
usually used for relative photometry in the optical and we list in
Table \ref{hostopt} also its flux. As Fig. \ref{oiii} shows, whereas
our approach did not succeed in bringing also the \OIII~$\lambda 5007$
line fluxes of the different epochs into alignment (bottom panel), it
did reduce the scatter of the original flux ratio distribution (middle
panel). Finally, for a subsample of 12 sources we compare in
Fig. \ref{reverb} the optical correction factors with the ratio
between the \OIII~$\lambda 5007$ line fluxes observed by reverberation
programmes \citep{Kor95, Pet98, Bentz06b, Denney10} and
us. Fig. \ref{reverb} indicates that the optical correction factors
are mostly overestimates relative to photometric conditions. This
result could be partly due to our near-IR flux-calibration, which was
based on the telluric standard stars. Since the target integration
times were considerably larger than those of the stars, the seeing
would have varied substantially for the former, leading overall to an
overestimate of the extraction aperture and so of the flux scale.


\begin{table*}
\caption{\label{hostir} 
Estimates of the Near-IR Host Galaxy Contribution}
\begin{tabular}{lcclllccccc}
\hline
Object Name & z & A$_{(1+z)5100}$ & host & ref. & IRTF & \SIII~$\lambda 9531$ & \multicolumn{2}{c}{near-IR spectrum} & {\sl HST} flux & {\sl HST} flux \\
&& [mag] & type && run & Flux & aperture & PA & (5580~\AA) & ($(1+z)5100$~\AA) \\
&&&&&& [erg/s/cm$^2$] & [arcsec] & [deg] & [erg/s/cm$^2$/\AA] & [erg/s/cm$^2$/\AA] \\
(1) & (2) & (3) & (4) & (5) & (6) & (7) & (8) & (9) & (10) & (11) \\
\hline
IRAS 1750$+$508 & 0.300 & 0.041 & ?    &     & 2004 May       & 9.88E$-$15 & 0.8 $\times$ 3.0 &   0 & ...        & 4.93E$-$17 \\
                &       &       &      &     & {\bf 2006 Jun} & 7.36E$-$15 & 0.8 $\times$ 4.0 &   0 & ...        & 6.06E$-$17 \\
H 1821$+$643    & 0.297 & 0.061 & E    & F04 & {\bf 2004 May} & 1.25E$-$14 & 0.8 $\times$ 3.2 &   0 & ...        & 5.29E$-$17 \\
PDS 456         & 0.184 & 0.941 & ?    &     & {\bf 2006 Jun} & 2.81E$-$15 & 0.8 $\times$ 4.0 &   0 & ...        & 1.19E$-$16 \\
3C 273          & 0.158 & 0.000 & E    & B09 & 2006 Jan       & 1.13E$-$14 & 0.8 $\times$ 5.4 & 207 & 2.61E$-$16 & 2.55E$-$16 \\
                &       &       &      &     & {\bf 2006 Jun} & 7.26E$-$15 & 0.8 $\times$ 4.0 &   0 & 2.28E$-$16 & 2.22E$-$16 \\
                &       &       &      &     & 2007 Jan       & 1.48E$-$14 & 0.8 $\times$ 5.4 &   0 & 2.50E$-$16 & 2.44E$-$16 \\
Mrk 876         & 0.129 & 0.005 & E    & B09 & {\bf 2006 Jun} & 1.17E$-$14 & 0.8 $\times$ 4.4 &   0 & 4.60E$-$16 & 4.54E$-$16 \\
HE 1228$+$013   & 0.117 & 0.000 & E    & L07 & {\bf 2006 Jun} & ...        & 0.8 $\times$ 3.4 &   0 & ...        & 1.75E$-$16 \\
                &       &       &      &     & 2007 Jan       & ...        & 0.8 $\times$ 5.4 &   0 & ...        & 2.41E$-$16 \\
PG 0844$+$349   & 0.064 & 0.031 & Sa   & B09 & 2006 Jan       & 4.57E$-$15 & 0.8 $\times$ 5.0 & 270 & 5.27E$-$16 & 5.00E$-$16 \\
                &       &       &      &     & {\bf 2007 Jan} & 7.10E$-$15 & 0.8 $\times$ 4.6 &   0 & 5.36E$-$16 & 5.08E$-$16 \\
Mrk 110         & 0.035 & 0.000 & Sa   & P07 & 2006 Jan       & 1.70E$-$14 & 0.8 $\times$ 5.4 & 254 & 2.49E$-$16 & 2.18E$-$16 \\
                &       &       &      &     & {\bf 2007 Jan} & 1.97E$-$14 & 0.8 $\times$ 5.0 &   0 & 2.47E$-$16 & 2.16E$-$16 \\
Mrk 509         & 0.034 & 0.083 & S0/a & P07 & 2004 May       & 3.23E$-$14 & 0.8 $\times$ 2.8 &   0 & 3.85E$-$16 & 3.81E$-$16 \\
                &       &       &      &     & {\bf 2006 Jun} & 4.25E$-$14 & 0.8 $\times$ 4.0 &   0 & 4.83E$-$16 & 4.77E$-$16 \\
Ark 120         & 0.033 & 0.570 & Sb   & P07 & 2006 Jan       & 2.61E$-$14 & 0.8 $\times$ 4.0 & 327 & 4.10E$-$15 & 5.87E$-$15 \\
                &       &       &      &     & {\bf 2007 Jan} & 2.28E$-$14 & 0.8 $\times$ 5.0 &   0 & 4.11E$-$15 & 5.88E$-$15 \\
Mrk 817         & 0.031 & 0.000 & Sa   & P07 & 2004 May       & 2.38E$-$14 & 0.8 $\times$ 3.6 &   0 & 2.26E$-$16 & 1.96E$-$16 \\
                &       &       &      &     & {\bf 2006 Jun} & 2.28E$-$14 & 0.8 $\times$ 4.0 &   0 & 2.51E$-$16 & 2.18E$-$16 \\
                &       &       &      &     & 2007 Jan       & 2.70E$-$14 & 0.8 $\times$ 5.0 &   0 & 2.95E$-$16 & 2.57E$-$16 \\
Mrk 290         & 0.030 & 0.000 & S0   & P07 & 2004 May       & 2.08E$-$14 & 0.8 $\times$ 3.0 &   0 & ...        & 4.71E$-$16 \\
                &       &       &      &     & {\bf 2006 Jun} & 3.33E$-$14 & 0.8 $\times$ 4.0 &   0 & ...        & 5.80E$-$16 \\
H 2106$-$099    & 0.027 & 0.161 & ?    &     & 2004 May       & 2.35E$-$14 & 0.8 $\times$ 3.6 &   0 & ...        & 5.74E$-$16 \\
                &       &       &      &     & {\bf 2006 Jun} & 2.25E$-$14 & 0.8 $\times$ 4.0 &   0 & ...        & 6.29E$-$16 \\
Mrk 335         & 0.026 & 0.077 & S0/a & P07 & 2006 Jan       & 1.31E$-$14 & 0.8 $\times$ 4.4 &  86 & 5.97E$-$16 & 5.44E$-$16 \\
                &       &       &      &     & {\bf 2007 Jan} & 1.28E$-$14 & 0.8 $\times$ 4.2 &   0 & 6.21E$-$16 & 5.67E$-$16 \\
Mrk 590         & 0.026 & 0.000 & S0   & P07 & 2006 Jan       & 1.10E$-$14 & 0.8 $\times$ 5.0 & 353 & 1.31E$-$15 & 1.11E$-$15 \\
                &       &       &      &     & {\bf 2007 Jan} & 3.29E$-$14 & 0.8 $\times$ 4.6 &   0 & 1.27E$-$15 & 1.08E$-$15 \\
Ark 564         & 0.025 & 0.228 & Sb   & O07 & {\bf 2006 Jun} & 3.22E$-$14 & 0.8 $\times$ 4.0 &   0 & ...        & 6.57E$-$16 \\
Mrk 79          & 0.022 & 0.183 & Sb   & P07 & 2006 Jan       & 5.09E$-$14 & 0.8 $\times$ 4.0 & 186 & 4.10E$-$16 & 4.06E$-$16 \\
                &       &       &      &     & {\bf 2007 Jan} & 4.87E$-$14 & 0.8 $\times$ 4.2 &   0 & 4.13E$-$16 & 4.08E$-$16 \\
NGC 5548        & 0.017 & 0.000 & Sa   & P07 & 2004 May       & 4.62E$-$14 & 0.8 $\times$ 5.0 &   0 & 1.22E$-$15 & 1.02E$-$15 \\
                &       &       &      &     & 2006 Jan       & 3.95E$-$14 & 0.8 $\times$ 8.0 & 273 & 1.38E$-$15 & 1.16E$-$15 \\
                &       &       &      &     & {\bf 2006 Jun} & 6.59E$-$14 & 0.8 $\times$ 4.0 &   0 & 1.14E$-$15 & 9.56E$-$16 \\
                &       &       &      &     & 2007 Jan       & 5.98E$-$14 & 0.8 $\times$ 5.8 &   0 & 1.26E$-$15 & 1.06E$-$15 \\
NGC 7469        & 0.016 & 0.140 & Sa   & P07 & {\bf 2006 Jan} & 1.16E$-$13 & 0.8 $\times$ 6.0 &  71 & 2.76E$-$15 & 2.59E$-$15 \\
H 1934$-$063    & 0.011 & 0.744 & ?    &     & {\bf 2006 Jun} & 1.14E$-$13 & 0.8 $\times$ 4.0 &   0 & ...        & 1.11E$-$15 \\
NGC 4593        & 0.009 & 0.000 & Sb   & P07 & 2004 May       & 2.56E$-$14 & 0.8 $\times$ 4.0 &   0 & 1.31E$-$15 & 1.08E$-$15 \\
                &       &       &      &     & {\bf 2006 Jun} & 2.89E$-$14 & 0.8 $\times$ 4.0 &   0 & 1.31E$-$15 & 1.08E$-$15 \\
                &       &       &      &     & 2007 Jan       & 3.76E$-$14 & 0.8 $\times$ 5.8 &   0 & 1.67E$-$15 & 1.38E$-$15 \\
NGC 3227        & 0.004 & 0.000 & Sa/b & RC3 & 2006 Jan       & 3.25E$-$13 & 0.8 $\times$ 6.0 &   0 & 2.22E$-$15 & 1.82E$-$15 \\
                &       &       &      &     & {\bf 2007 Jan} & 2.68E$-$13 & 0.8 $\times$ 5.0 &   0 & 2.17E$-$15 & 1.77E$-$15 \\
NGC 4151        & 0.003 & 0.000 & Sa/b & RC3 & 2004 May       & 1.06E$-$12 & 0.8 $\times$ 7.6 &   0 & 5.71E$-$15 & 4.66E$-$15 \\
                &       &       &      &     & 2006 Jan       & 1.15E$-$12 & 0.8 $\times$ 6.0 & 218 & 5.79E$-$15 & 4.72E$-$15 \\
                &       &       &      &     & {\bf 2006 Jun} & 1.16E$-$12 & 0.8 $\times$ 4.4 &   0 & 5.03E$-$15 & 4.10E$-$15 \\
                &       &       &      &     & 2007 Jan       & 8.22E$-$13 & 0.8 $\times$ 5.8 &   0 & 5.39E$-$15 & 4.40E$-$15 \\ 
\hline
\end{tabular}

\parbox[]{18.7cm}{The columns are: (1) object name; (2) redshift from
  the NASA/IPAC Extragalactic Database (NED); (3) Galactic extinction
  at rest-frame 5100~\AA; (4) Hubble type of the host galaxy; (5) 
  reference for the host type, where P07: \citet{Petr07}, B09:
  \citet{Bentz09}, RC3: \citet{RC3}, F04: \citet{Floyd04}, L07:
  \citet{Let07}, O07: \citet{Ohta07}; (6) IRTF run (epoch with best
  weather conditions in bold); (7) observed flux of the narrow line 
  \SIII~$\lambda 9531$; (8) near-IR extraction aperture; (9) near-IR 
  slit position angle, where PA=0$^{\circ}$ corresponds to E-W orientation
  and is defined E through N; (10) host galaxy flux at 5580~\AA~(except 
  for Mrk 509 at 5483~\AA) estimated from {\sl HST} images of 
  \citet{Bentz06a, Bentz09}; and (11) host galaxy flux at rest-frame
  5100~\AA, corrected for Galactic extinction using the values in column (3).}

\end{table*}

\begin{table*}
\caption{\label{hostopt} 
Estimates of the Optical Host Galaxy Contribution}
\begin{tabular}{lcclcccrccc}
\hline
Object Name & z & A$_{(1+z)5100}$ & IRTF & \OIII~$\lambda 5007$ & opt. & \multicolumn{2}{c}{optical spectrum} & 
{\sl HST} flux & {\sl HST} flux & host \\
&& [mag] & run & Flux & corr. & aperture & PA & (5580~\AA) & ($(1+z)5100$~\AA) & corr. \\
&&&& [erg/s/cm$^2$] & factor & [arcsec] & [deg] & [erg/s/cm$^2$/\AA] & [erg/s/cm$^2$/\AA] & factor \\
(1) & (2) & (3) & (4) & (5) & (6) & (7) & (8) & (9) & (10) & (11) \\
\hline
IRAS 1750$+$508 & 0.300 & 0.041 & 2004 May & 8.43E$-$14 & 0.90 & 3 $\times$ 4.2 &   110 & ...        & 1.67E$-$16 & 1.0 \\
                &       &       & 2006 Jun & 7.64E$-$14 & 1.00 & 3 $\times$ 4.8 &    90 & ...        & 1.83E$-$16 & 1.0 \\
H 1821$+$643    & 0.297 & 0.061 & 2004 May & 1.63E$-$13 & 1.32 & 3 $\times$ 3.6 &   110 & ...        & 1.49E$-$16 & 1.0 \\
PDS 456         & 0.184 & 0.941 & 2006 Jun & 1.22E$-$14 & 1.70 & 3 $\times$ 4.2 &     6 & ...        & 3.20E$-$16 & 1.0 \\
3C 273          & 0.158 & 0.000 & 2006 Jan & 1.92E$-$13 & 0.67 & 3 $\times$ 3.6 & $-$15 & 5.49E$-$16 & 5.36E$-$16 & 1.0 \\
                &       &       & 2006 Jun & 1.51E$-$13 & 1.80 & 3 $\times$ 4.2 &    90 & 6.00E$-$16 & 5.86E$-$16 & 1.0 \\
                &       &       & 2007 Jan & 2.11E$-$13 & 0.75 & 3 $\times$ 4.8 &    25 & 6.27E$-$16 & 6.12E$-$16 & 1.0 \\
Mrk 876         & 0.129 & 0.005 & 2006 Jun & 5.13E$-$14 & 1.40 & 3 $\times$ 4.8 &   110 & 7.93E$-$16 & 7.83E$-$16 & 1.0 \\
HE 1228$+$013   & 0.117 & 0.000 & 2006 Jun & 4.44E$-$14 & 1.50 & 3 $\times$ 4.8 &    50 & ...        & 5.78E$-$16 & 1.0 \\
                &       &       & 2007 Jan & 5.61E$-$14 & 1.80 & 3 $\times$ 4.2 &    33 & ...        & 5.27E$-$16 & 1.0 \\
PG 0844$+$349   & 0.064 & 0.031 & 2006 Jan & 5.24E$-$14 & 1.80 & 3 $\times$ 3.6 &    80 & 7.61E$-$16 & 7.22E$-$16 & 4.9 \\
                &       &       & 2007 Jan & 4.56E$-$14 & 1.75 & 3 $\times$ 7.8 &   107 & 8.96E$-$16 & 8.51E$-$16 & 3.6 \\
Mrk 110         & 0.035 & 0.000 & 2006 Jan & 2.00E$-$13 & 2.60 & 3 $\times$ 4.8 &     6 & 5.55E$-$16 & 4.86E$-$16 & 5.4 \\
                &       &       & 2007 Jan & 2.66E$-$13 & 2.10 & 3 $\times$ 6.0 & $-$15 & 5.86E$-$16 & 5.13E$-$16 & 6.5 \\
Mrk 509         & 0.034 & 0.083 & 2004 May & 4.63E$-$13 & 2.80 & 3 $\times$ 4.2 &     0 & 1.49E$-$15 & 1.47E$-$15 & 4.2 \\
                &       &       & 2006 Jun & 6.11E$-$13 & 2.30 & 3 $\times$ 4.8 & $-$10 & 1.62E$-$15 & 1.60E$-$15 & 4.5 \\
Ark 120         & 0.033 & 0.570 & 2006 Jan & 1.94E$-$13 & 1.00 & 3 $\times$ 4.8 &     7 & 5.78E$-$15 & 8.28E$-$15 & 1.0 \\
                &       &       & 2007 Jan & 1.69E$-$13 & 1.00 & 3 $\times$ 4.8 &     6 & 5.78E$-$15 & 8.28E$-$15 & 1.0 \\
Mrk 817         & 0.031 & 0.000 & 2004 May & 3.81E$-$13 & 1.00 & 3 $\times$ 4.2 &     0 & 8.60E$-$16 & 7.48E$-$16 & 2.9 \\
                &       &       & 2006 Jun & 1.16E$-$13 & 1.90 & 3 $\times$ 4.2 &   110 & 8.65E$-$16 & 7.53E$-$16 & 1.7 \\
                &       &       & 2007 Jan & 8.28E$-$14 & 3.10 & 3 $\times$ 7.2 &    20 & 1.11E$-$15 & 9.66E$-$16 & 3.3 \\
Mrk 290         & 0.030 & 0.000 & 2004 May & 2.10E$-$13 & 3.40 & 3 $\times$ 4.8 &   110 & ...        & 1.75E$-$15 & 2.0 \\
                &       &       & 2006 Jun & 1.93E$-$13 & 3.30 & 3 $\times$ 5.4 &    90 & ...        & 1.92E$-$15 & 1.0 \\
H 2106$-$099    & 0.027 & 0.161 & 2004 May & 1.23E$-$13 & 2.75 & 3 $\times$ 4.2 &  $-$5 & ...        & 1.69E$-$15 & 2.5 \\
                &       &       & 2006 Jun & 1.52E$-$13 & 2.00 & 3 $\times$ 5.4 & $-$14 & ...        & 2.04E$-$15 & 2.1 \\
Mrk 335         & 0.026 & 0.077 & 2006 Jan & 2.78E$-$13 & 1.00 & 3 $\times$ 3.6 &    55 & 1.20E$-$15 & 1.09E$-$15 & 1.7 \\
                &       &       & 2007 Jan & 1.76E$-$13 & 1.90 & 3 $\times$ 6.6 &    65 & 1.45E$-$15 & 1.32E$-$15 & 2.4 \\
Mrk 590         & 0.026 & 0.000 & 2006 Jan & 8.18E$-$14 & 1.00 & 3 $\times$ 6.6 &     5 & 3.65E$-$15 & 3.11E$-$15 & 1.5 \\
                &       &       & 2007 Jan & 8.09E$-$14 & 1.00 & 3 $\times$ 9.6 &    11 & 3.96E$-$15 & 3.37E$-$15 & 1.0 \\
Ark 564         & 0.025 & 0.228 & 2006 Jun & 2.25E$-$13 & 1.75 & 3 $\times$ 4.8 &    70 & ...        & 1.98E$-$15 & 1.3 \\
Mrk 79          & 0.022 & 0.183 & 2006 Jan & 4.79E$-$13 & 1.60 & 3 $\times$ 4.2 &    28 & 1.03E$-$15 & 1.02E$-$15 & 2.8 \\
                &       &       & 2007 Jan & 4.58E$-$13 & 1.75 & 3 $\times$ 7.8 & $-$45 & 1.27E$-$15 & 1.27E$-$15 & 4.1 \\
NGC 5548        & 0.017 & 0.000 & 2004 May & 1.58E$-$12 & 0.55 & 3 $\times$ 7.2 &    71 & 3.37E$-$15 & 2.82E$-$15 & 1.0 \\
                &       &       & 2006 Jan & 5.90E$-$13 & 2.40 & 3 $\times$ 6.0 &   110 & 3.24E$-$15 & 2.71E$-$15 & 2.1 \\
                &       &       & 2006 Jun & 4.35E$-$13 & 2.10 & 3 $\times$ 8.4 &    70 & 3.54E$-$15 & 2.96E$-$15 & 1.4 \\
                &       &       & 2007 Jan & 3.06E$-$13 & 2.50 & 3 $\times$ 8.4 & $-$55 & 3.58E$-$15 & 2.99E$-$15 & 2.0 \\
NGC 7469        & 0.016 & 0.140 & 2006 Jan & 9.95E$-$13 & 1.00 & 3 $\times$ 4.8 &    51 & 7.89E$-$15 & 7.41E$-$15 & 1.0 \\
H 1934$-$063    & 0.011 & 0.744 & 2006 Jun & 5.52E$-$13 & 2.10 & 3 $\times$ 5.4 & $-$26 & ...        & 3.59E$-$15 & 1.8 \\
NGC 4593        & 0.009 & 0.000 & 2004 May & 1.74E$-$13 & 2.25 & 3 $\times$ 6.0 &     5 & 4.81E$-$15 & 3.97E$-$15 & 3.1 \\
                &       &       & 2006 Jun & 1.54E$-$13 & 1.85 & 3 $\times$ 7.2 &    43 & 5.39E$-$15 & 4.44E$-$15 & 1.9 \\
                &       &       & 2007 Jan & 1.13E$-$13 & 3.80 & 3 $\times$ 9.0 &    10 & 5.64E$-$15 & 4.65E$-$15 & 4.9 \\
NGC 3227        & 0.004 & 0.000 & 2006 Jan & 9.10E$-$13 & 1.00 & 3 $\times$ 6.0 &     0 & 6.46E$-$15 & 5.28E$-$15 & 1.5 \\
                &       &       & 2007 Jan & 1.28E$-$12 & 0.71 & 3 $\times$10.2 & $-$13 & 7.62E$-$15 & 6.23E$-$15 & 1.5 \\
NGC 4151        & 0.003 & 0.000 & 2004 May & 8.03E$-$12 & 3.00 & 3 $\times$ 5.4 & $-$40 & 1.22E$-$14 & 9.91E$-$15 & 3.7 \\
                &       &       & 2006 Jan & 1.18E$-$11 & 1.30 & 3 $\times$ 4.8 &    54 & 1.19E$-$14 & 9.71E$-$15 & 2.2 \\
                &       &       & 2006 Jun & 1.18E$-$11 & 2.00 & 3 $\times$ 6.6 &    90 & 1.33E$-$14 & 1.08E$-$14 & 1.8 \\
                &       &       & 2007 Jan &$>$4.09E$-$12&2.40 & 3 $\times$ 8.4 &   102 & 1.42E$-$14 & 1.15E$-$14 & 3.9 \\ 
\hline
\end{tabular}

\parbox[]{18.75cm}{The columns are: (1) object name; (2) redshift from
  the NASA/IPAC Extragalactic Database (NED); (3) Galactic extinction
  at rest-frame 5100~\AA; (4) IRTF run; (5) observed flux of the narrow-line 
  \OIII~$\lambda 5007$; (6) flux scale correction factor for optical spectrum; 
  (7) optical extraction aperture; (8) optical slit position angle, where 
  PA=90$^{\circ}$ corresponds to E-W orientation; (9) observed host galaxy
  flux at 5580~\AA~(except for Mrk 509 at 5483~\AA) in optical aperture derived
  from {\sl HST} images of \citet{Bentz06a, Bentz09}; (10) observed host galaxy
  flux at rest-frame 5100~\AA~in optical aperture, corrected for Galactic extinction
  using the values in column (3); and (11) host galaxy flux correction factor.}

\end{table*}

\section{The AGN continuum components}


\begin{figure*}
\centerline{
\includegraphics[scale=0.95]{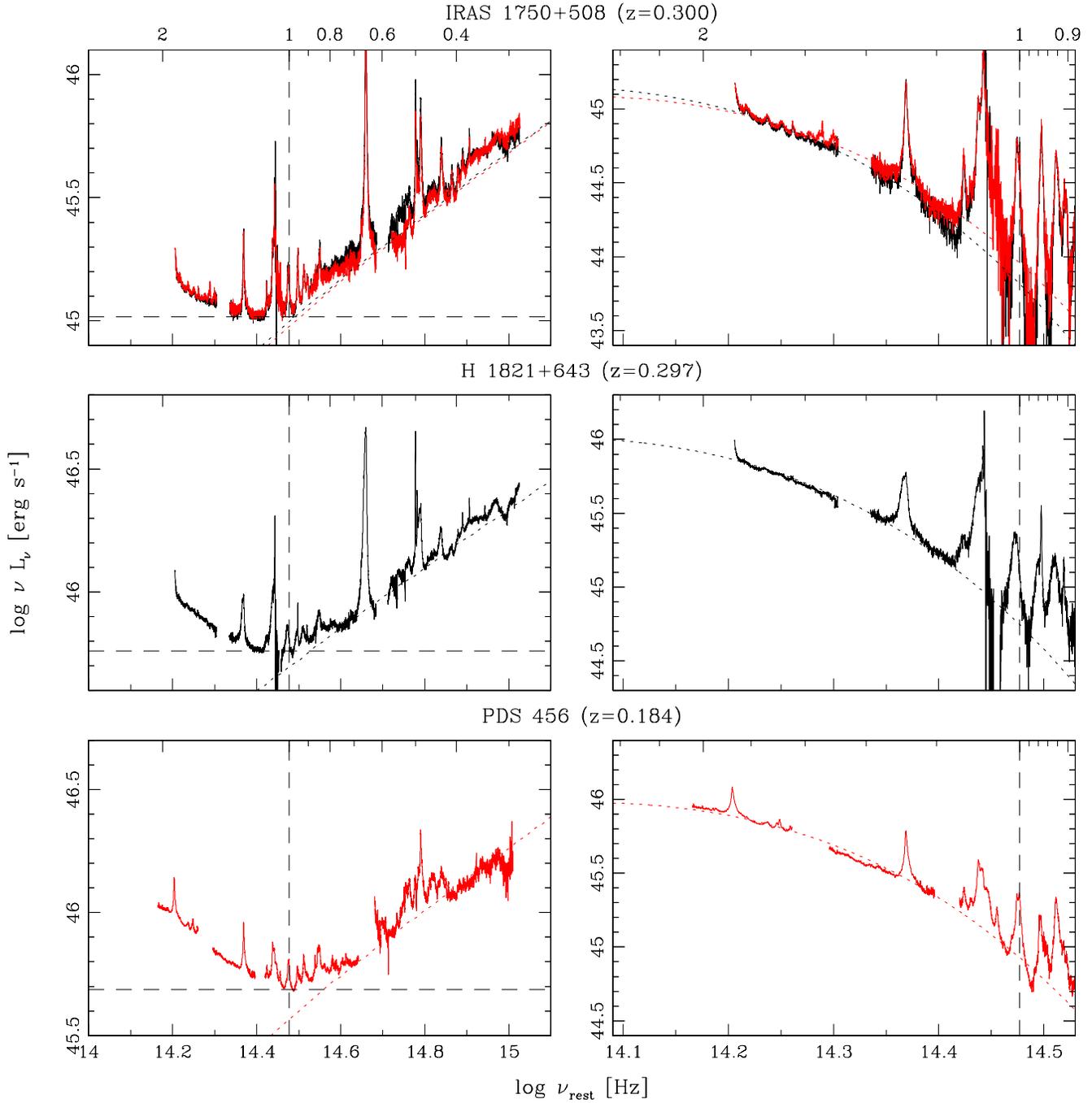}
}
\caption{\label{SEDfit} Left panels: Rest-frame AGN (host
  galaxy-subtracted) spectral energy distributions (from right panels
  of Fig. \ref{SEDsub}) normalized at 1 $\mu$m (vertical dashed line)
  to the luminosity of the lowest-flux epoch. The four IRTF observing
  runs are shown in colour: 2004 May (black), 2006 January (green),
  2006 June (red), and 2007 January (blue). The dotted line indicates
  the accretion disc spectrum that approximates best the continuum
  blueward of $\sim 1$ $\mu$m. The horizontal dashed line marks the
  minimum in the integrated luminosity of the lowest-flux
  epoch. Wavelength units in $\mu$m are labeled on the top axis. Right
  panels: As in left panels with the accretion disc spectrum
  subtracted. The dotted line shows the blackbody spectrum fitted to
  the continuum redward of $\sim 1$ $\mu$m.}
\end{figure*}

\setcounter{figure}{5}
\begin{figure*}
\centerline{
\includegraphics[scale=0.95]{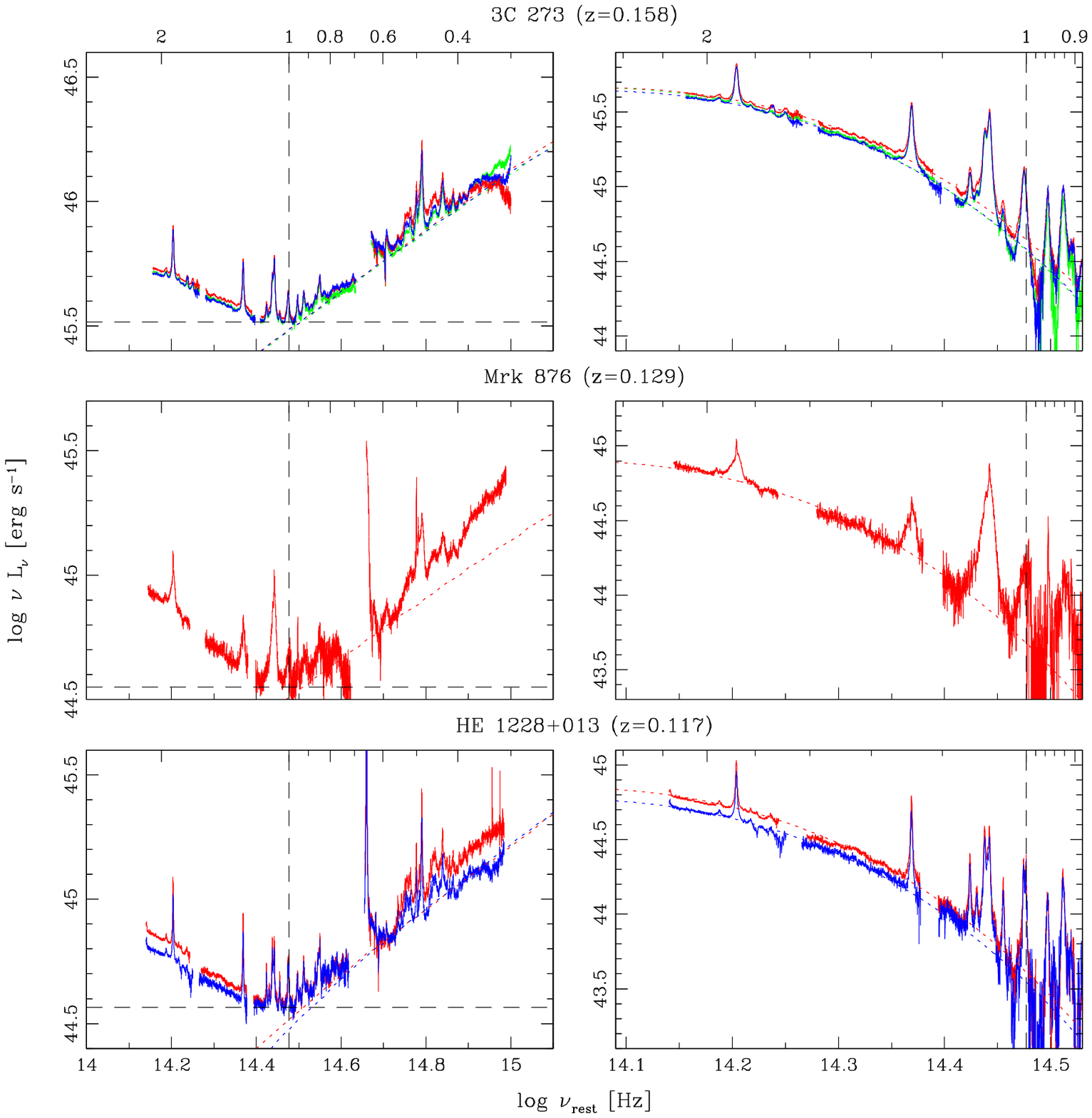}
}
\contcaption{}
\end{figure*}

\setcounter{figure}{5}
\begin{figure*}
\centerline{
\includegraphics[scale=0.95]{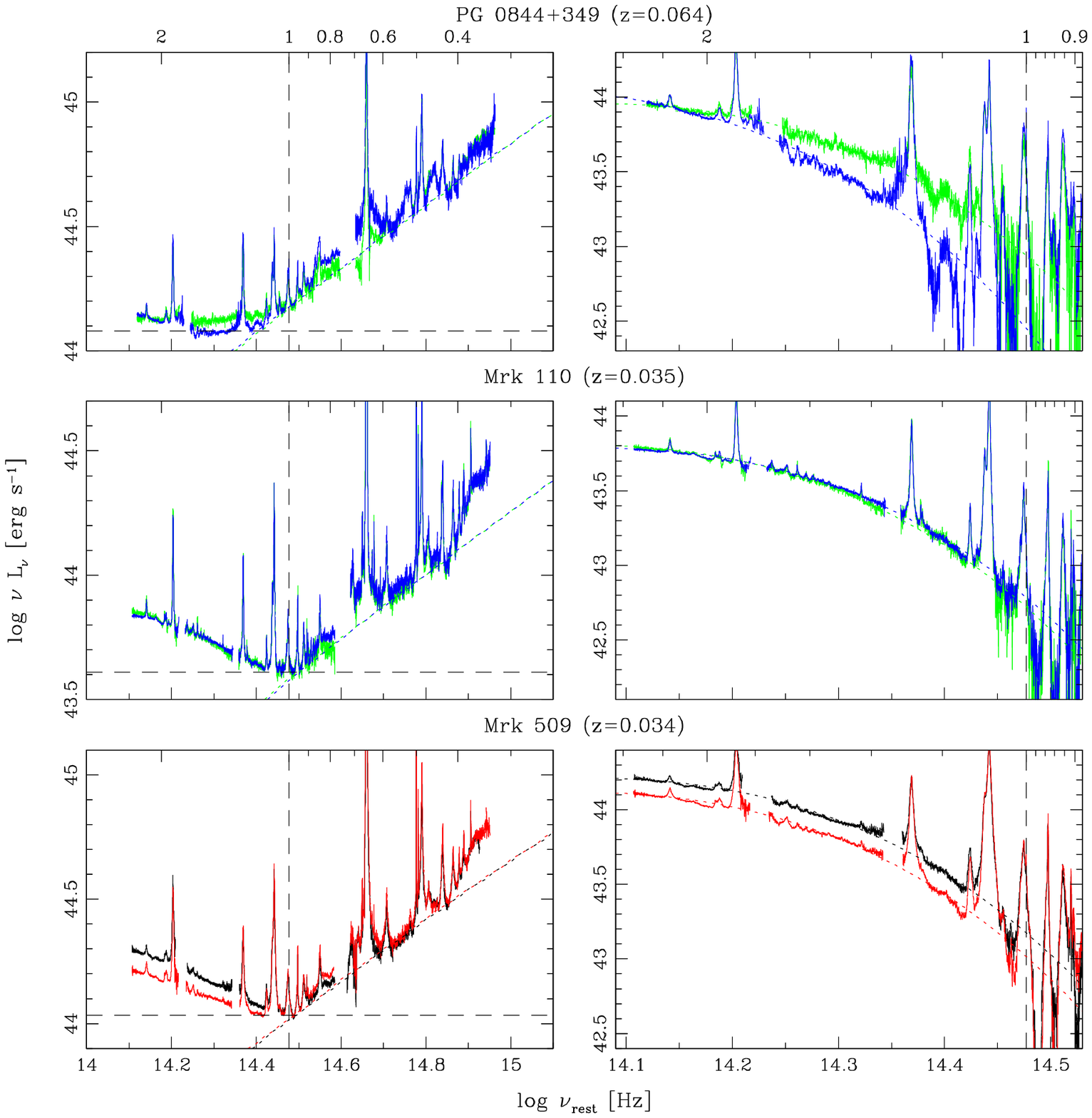}
}
\contcaption{}
\end{figure*}

\setcounter{figure}{5}
\begin{figure*}
\centerline{
\includegraphics[scale=0.95]{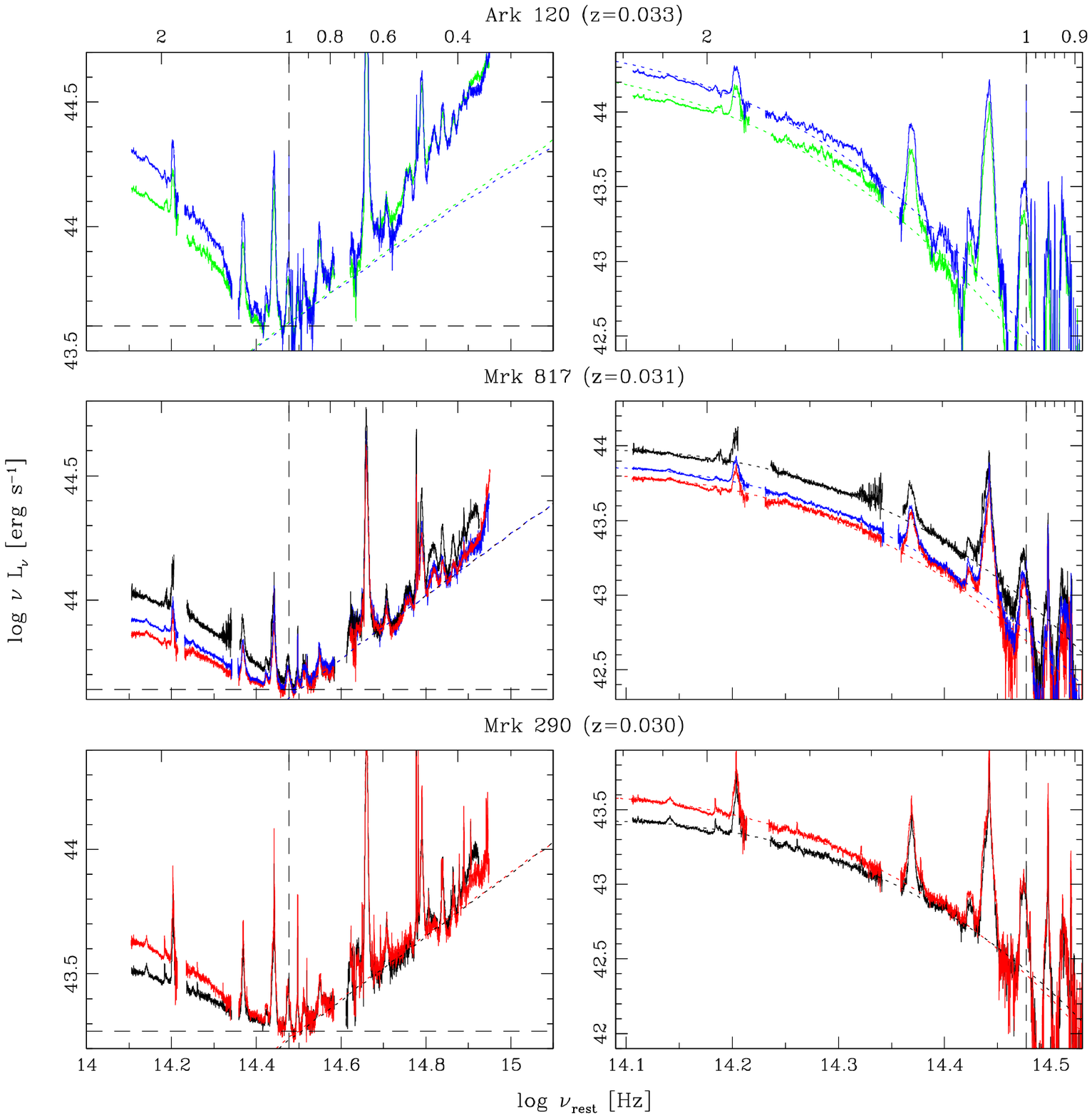}
}
\contcaption{}
\end{figure*}

\setcounter{figure}{5}
\begin{figure*}
\centerline{
\includegraphics[scale=0.95]{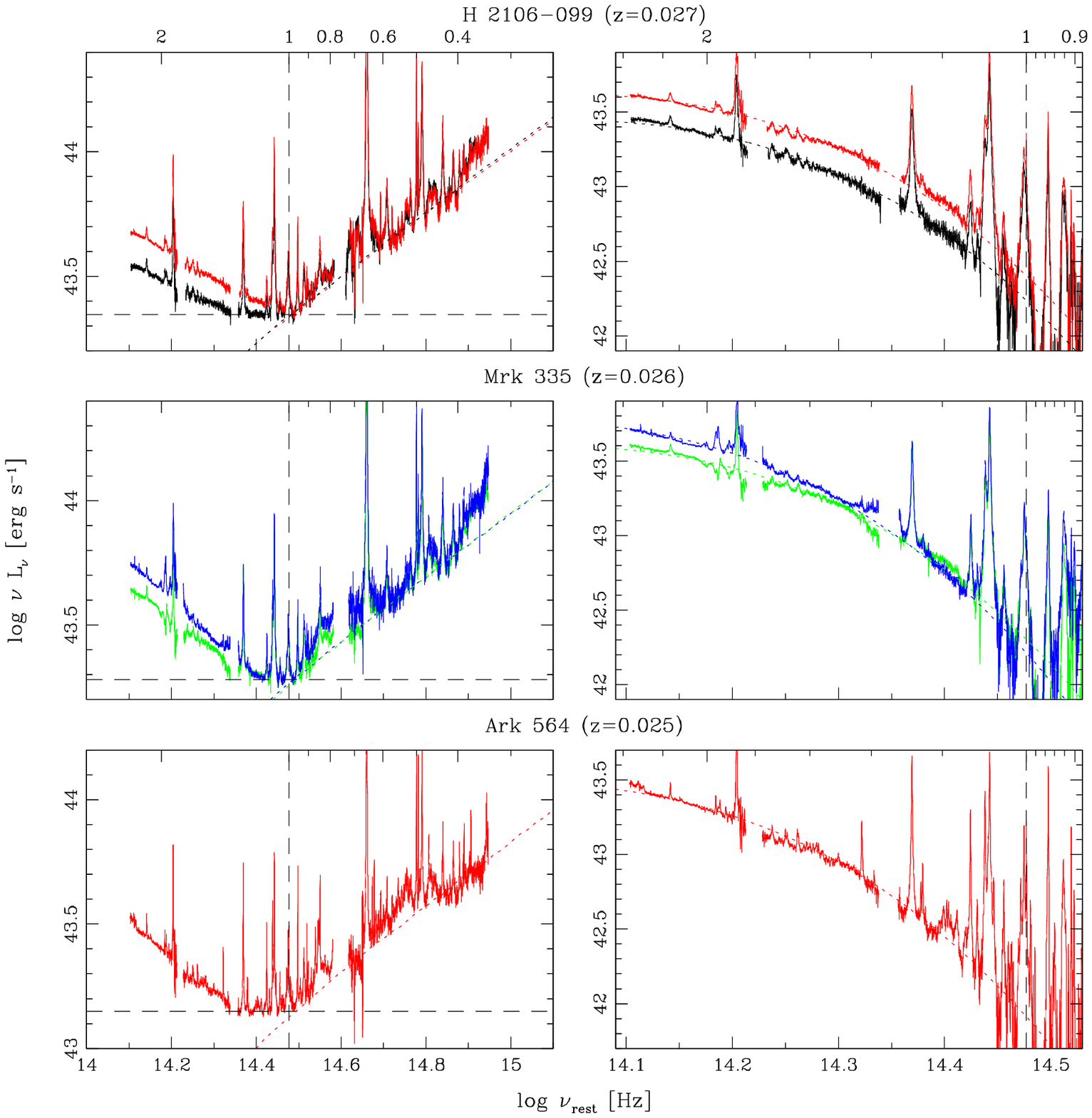}
}
\contcaption{}
\end{figure*}

\setcounter{figure}{5}
\begin{figure*}
\centerline{
\includegraphics[scale=0.95]{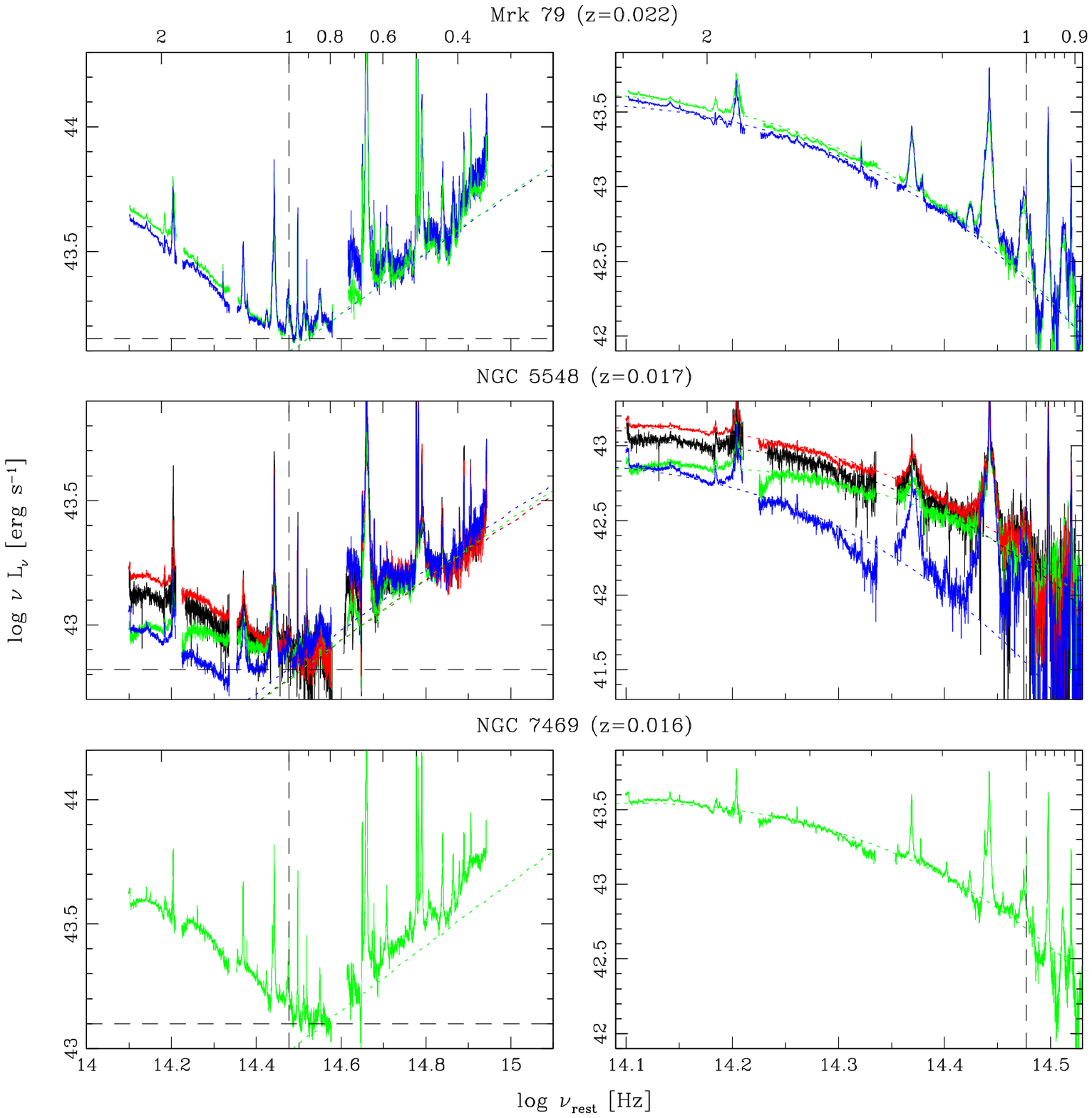}
}
\contcaption{}
\end{figure*}

\setcounter{figure}{5}
\begin{figure*}
\centerline{
\includegraphics[scale=0.95]{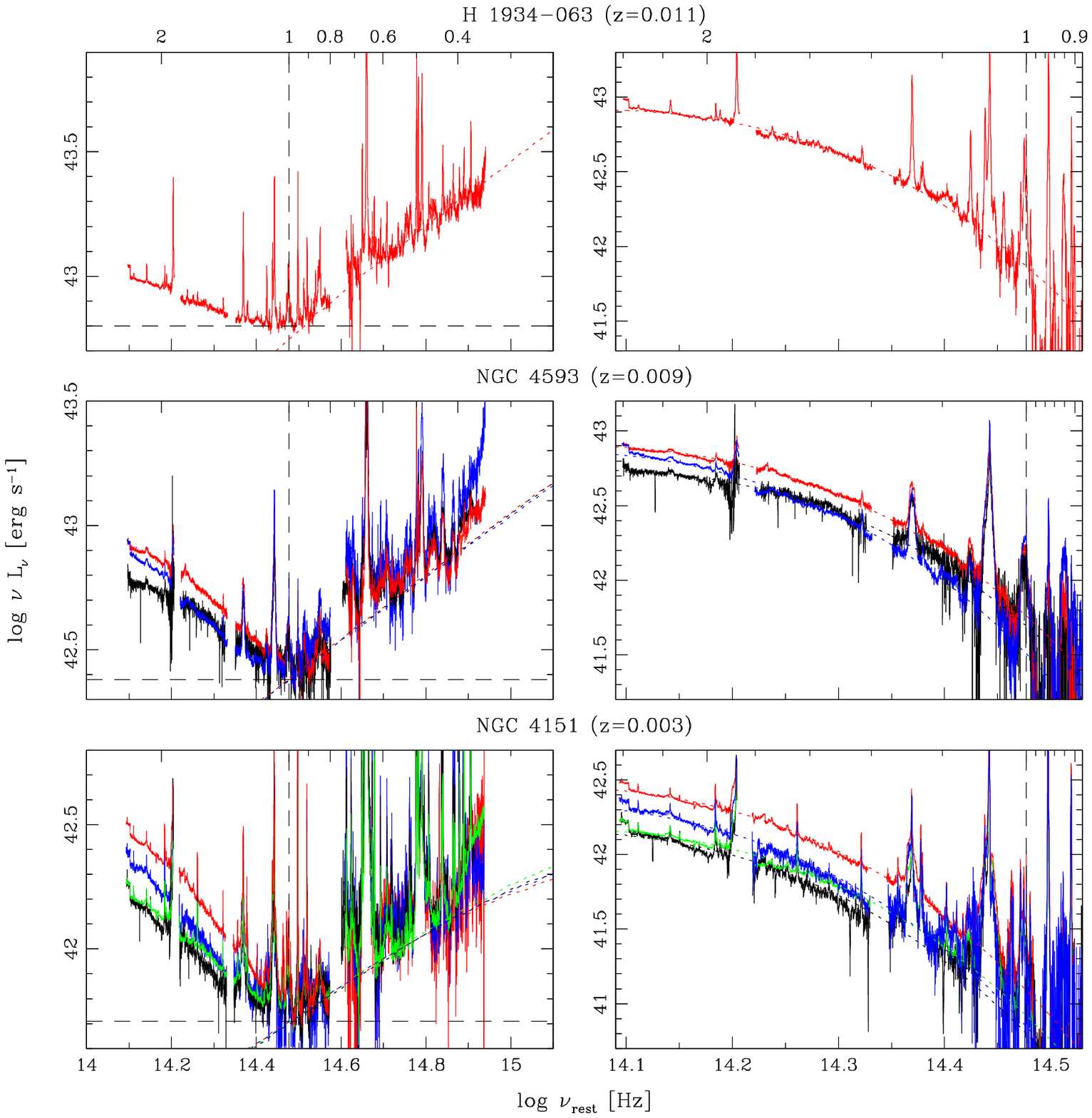}
}
\contcaption{}
\end{figure*}

\begin{figure*}
\centerline{
\includegraphics[bb=18 506 591 718,clip=true,scale=1.0]{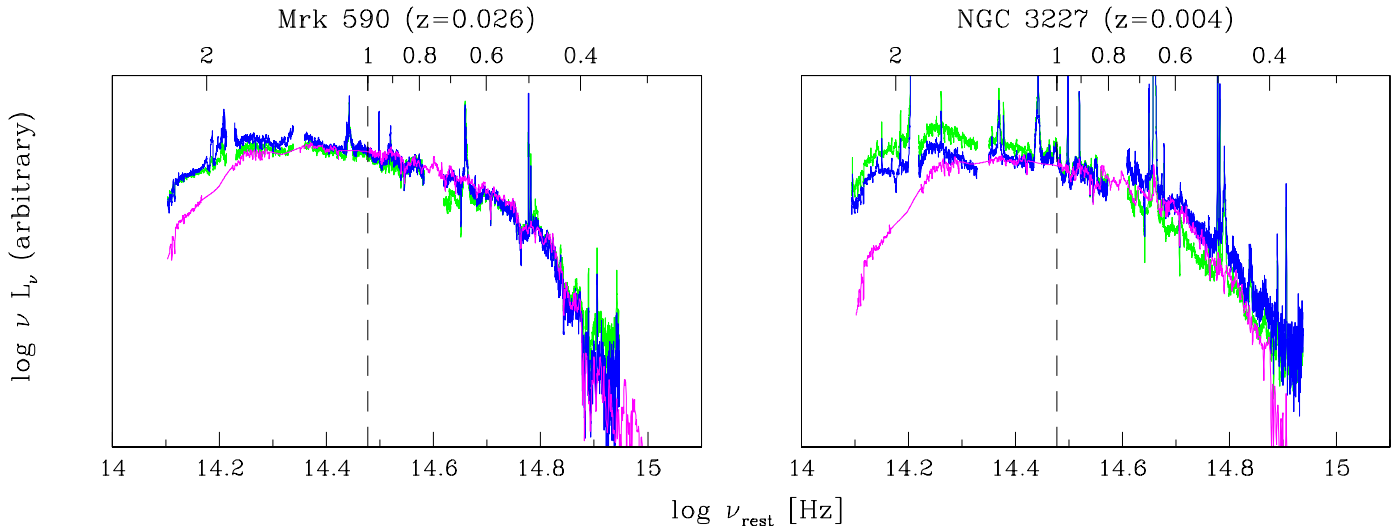}
}
\caption{\label{SEDgal} Rest-frame spectral energy distributions for
  the two sources (Mrk 590 and NGC 3227) that are strongly affected by
  host galaxy light. The spectra corresponding to the IRTF observing
  runs in 2006 January (green) and 2007 January (blue) are normalized
  at 1 $\mu$m (dashed line). The host galaxy templates from
  \citet{Man01} are overplotted (magenta). Wavelength units in $\mu$m
  are labeled on the top axis.}
\end{figure*}


\begin{table*}
\caption{\label{fitstab} 
AGN Continuum Fit Parameters}
\begin{tabular}{lcclccccccc}
\hline
Object Name & $M_{\rm BH}$ & Ref. & IRTF run & log $\nu L_{1\mu m}$ & var. & 
log $\nu L_{\rm acc}$ & $T_{\rm hot}$ & log $\nu L_{\rm hot}$ & $C$ & $R_{\rm hot}$ \\
& [$M_{\odot}$] &&& [erg s$^{-1}$] & factor & [erg s$^{-1}$] & [K] & [erg s$^{-1}$] & & [lyr] \\ 
(1) & (2) & (3) & (4) & (5) & (6) & (7) & (8) & (9) & (10) & (11) \\
\hline
IRAS 1750$+$508 & 9.3e$+$07 & est & 2004 May & 45.02 & 1.00 & 46.84 & 1289 & 45.15 & 0.01 & 6.27 \\
                &           &     & 2006 Jun & 45.12 & 1.26 & 46.98 & 1383 & 45.19 & 0.01 & 6.40 \\
H 1821$+$643    & 8.9e$+$08 & est & 2004 May & 45.76 & 1.00 & 46.92 & 1320 & 46.01 & 0.05 & 6.55 \\
PDS 456         & 1.8e$+$08 & est & 2006 Jun & 45.69 & 1.00 & 47.42 & 1425 & 45.98 & 0.01 &10.00 \\
3C 273          & 8.9e$+$08 & P04 & 2006 Jan & 45.56 & 1.10 & 46.66 & 1403 & 45.71 & 0.04 & 4.30 \\
                &           &     & 2006 Jun & 45.84 & 1.91 & 47.02 & 1443 & 45.94 & 0.03 & 6.15 \\
                &           &     & 2007 Jan & 45.52 & 1.00 & 46.61 & 1418 & 45.64 & 0.04 & 3.97 \\
Mrk 876         & 2.8e$+$08 & P04 & 2006 Jun & 44.55 & 1.00 & 45.65 & 1339 & 44.90 & 0.07 & 1.48 \\
HE 1228$+$013   & 5.2e$+$07 & est & 2006 Jun & 44.57 & 1.00 & 46.38 & 1333 & 44.85 & 0.01 & 3.45 \\
                &           &     & 2007 Jan & 44.95 & 2.40 & 46.95 & 1337 & 45.15 & 0.01 & 6.61 \\
PG 0844$+$349   & 9.2e$+$07 & P04 & 2006 Jan & 44.38 & 1.58 & 45.89 & 1443 & 44.16 & 0.01 & 1.68 \\
                &           &     & 2007 Jan & 44.18 & 1.00 & 45.61 & 1190 & 44.05 & 0.01 & 1.78 \\
Mrk 110         & 2.5e$+$07 & P04 & 2006 Jan & 43.61 & 1.00 & 45.28 & 1406 & 43.81 & 0.01 & 0.87 \\
                &           &     & 2007 Jan & 43.71 & 1.26 & 45.42 & 1452 & 43.89 & 0.01 & 0.96 \\
Mrk 509         & 1.4e$+$08 & P04 & 2004 May & 44.11 & 1.20 & 45.31 & 1432 & 44.29 & 0.04 & 0.87 \\
                &           &     & 2006 Jun & 44.03 & 1.00 & 45.20 & 1398 & 44.12 & 0.03 & 0.81 \\
Ark 120         & 1.5e$+$08 & P04 & 2006 Jan & 43.80 & 1.58 & 44.88 & 1102 & 44.49 & 0.16 & 0.90 \\
                &           &     & 2007 Jan & 43.60 & 1.00 & 44.58 & 1100 & 44.43 & 0.28 & 0.64 \\
Mrk 817         & 4.3e$+$07 & D10 & 2004 May & 43.64 & 1.00 & 45.08 & 1446 & 43.97 & 0.03 & 0.66 \\
                &           &     & 2006 Jun & 43.64 & 1.00 & 45.08 & 1392 & 43.81 & 0.02 & 0.71 \\
                &           &     & 2007 Jan & 43.64 & 1.00 & 45.08 & 1401 & 43.86 & 0.02 & 0.70 \\
Mrk 290         & 2.4e$+$07 & D10 & 2004 May & 43.47 & 1.58 & 45.07 & 1453 & 43.62 & 0.01 & 0.64 \\
                &           &     & 2006 Jun & 43.27 & 1.00 & 44.78 & 1353 & 43.59 & 0.03 & 0.53 \\
H 2106$-$099    & 2.3e$+$07 & est & 2004 May & 43.45 & 1.26 & 45.10 & 1346 & 43.55 & 0.01 & 0.78 \\
                &           &     & 2006 Jun & 43.35 & 1.00 & 44.95 & 1342 & 43.62 & 0.02 & 0.66 \\
Mrk 335         & 1.4e$+$07 & P04 & 2006 Jan & 43.42 & 1.38 & 45.26 & 1308 & 43.74 & 0.01 & 0.99 \\
                &           &     & 2007 Jan & 43.28 & 1.00 & 45.06 & 1206 & 43.77 & 0.02 & 0.92 \\
Ark 564         & 9.8e$+$06 & est & 2006 Jun & 43.15 & 1.00 & 45.03 & 1202 & 43.48 & 0.01 & 0.90 \\
Mrk 79          & 5.2e$+$07 & P04 & 2006 Jan & 43.24 & 1.23 & 44.38 & 1339 & 43.71 & 0.09 & 0.34 \\
                &           &     & 2007 Jan & 43.15 & 1.00 & 44.25 & 1364 & 43.55 & 0.08 & 0.28 \\
NGC 5548        & 4.4e$+$07 & D10 & 2004 May & 42.82 & 1.00 & 43.87 & 1572 & 43.03 & 0.06 & 0.14 \\
                &           &     & 2006 Jan & 43.22 & 2.51 & 44.44 & 1730 & 43.26 & 0.03 & 0.22 \\
                &           &     & 2006 Jun & 42.92 & 1.26 & 44.01 & 1547 & 43.22 & 0.06 & 0.17 \\
                &           &     & 2007 Jan & 42.92 & 1.26 & 44.07 & 1291 & 42.98 & 0.03 & 0.26 \\
NGC 7469        & 1.2e$+$07 & P04 & 2006 Jan & 43.10 & 1.00 & 44.70 & 1551 & 43.54 & 0.03 & 0.37 \\
H 1934$-$063    & 5.8e$+$06 & est & 2006 Jun & 42.80 & 1.00 & 44.70 & 1426 & 42.91 & 0.01 & 0.44 \\
NGC 4593        & 9.8e$+$06 & D06 & 2004 May & 42.58 & 1.58 & 44.18 & 1429 & 42.94 & 0.02 & 0.24 \\
                &           &     & 2006 Jun & 42.55 & 1.48 & 44.14 & 1380 & 43.07 & 0.03 & 0.24 \\
                &           &     & 2007 Jan & 42.38 & 1.00 & 43.88 & 1281 & 42.87 & 0.04 & 0.21 \\
NGC 4151        & 4.6e$+$07 & B06 & 2004 May & 41.88 & 1.48 & 42.53 & 1281 & 42.33 & 0.25 & 0.04 \\
                &           &     & 2006 Jan & 41.97 & 1.82 & 42.67 & 1328 & 42.43 & 0.23 & 0.05 \\
                &           &     & 2006 Jun & 41.71 & 1.00 & 42.31 & 1278 & 42.45 & 0.55 & 0.03 \\
                &           &     & 2007 Jan & 41.81 & 1.26 & 42.44 & 1231 & 42.44 & 0.40 & 0.04 \\ 
\hline
\end{tabular}

\parbox[]{15.7cm}{The columns are: (1) object name; (2) black hole mass (in solar masses); 
  (3) reference for the black hole mass, where B06: \citet{Bentz06b}, D06: \citet{Denney06}, 
  D10: \citet{Denney10}, P04: \citet{Pet04} and est: estimated based on the correlation 
  in Fig. \ref{bhmass}; (4) IRTF run; (5) total AGN (host galaxy-subtracted) luminosity 
  at 1 $\mu$m; (6) variability factor relative to the lowest-flux epoch; (7) peak luminosity 
  of the accretion disc; for the hot dust component (8) its blackbody temperature, (9) peak 
  luminosity, (10) covering factor, and (11) average radius (in light years).}

\end{table*}


Our observations cover the continuum rest-frame frequency range of
$\nu \sim 10^{14} - 10^{15}$ Hz, which is believed to sample two
important emission components, namely, the accretion disc
\citep[e.g.,][]{Mal82, Mal83} and the hottest part of the putative
dusty torus \citep[e.g.,][]{Barv87, Neu87}. These two components are
predicted to have opposite spectral behaviours, namely, the accretion
disc and the hot dust emission rising and falling towards longer
frequencies (in a logarithmic $\nu f_\nu$ versus $\nu$ plot),
respectively, resulting in an inflection point at the location where
they meet \citep[at $\sim 1$ $\mu$m; e.g.,][]{Ward87a, Carl87,
  Elvis94, Glik06, Rif06}.

Our present knowledge of this special continuum region is based mainly
on optical spectroscopy combined with near-IR photometry, with the two
frequency ranges usually not observed contemporaneously. Our
observations now give us for the first time the opportunity to define
the separate continuum components at spectroscopic rather than
photometric precision. This permits us to investigate their properties
in greater detail than was possible before. In particular, we are
interested to understand if a near-IR reverberation mapping campaign
generally requires simultaneous optical spectroscopy to determine the
state of the ionising flux or if the latter can also be derived from
the continuum behaviour at longer wavelengths.

For the purpose of the following analysis we have normalised all SEDs
at $\sim 1$ $\mu$m (Fig. \ref{SEDfit}). The SEDs of two sources,
namely, Mrk~590 and NGC~3227, are strongly affected by host galaxy
light in both their near-IR and optical spectral parts
(Fig. \ref{SEDgal}) and are not considered further.

\subsection{The accretion disc} \label{accretion}

\begin{figure}
\centerline{
\includegraphics[scale=0.4]{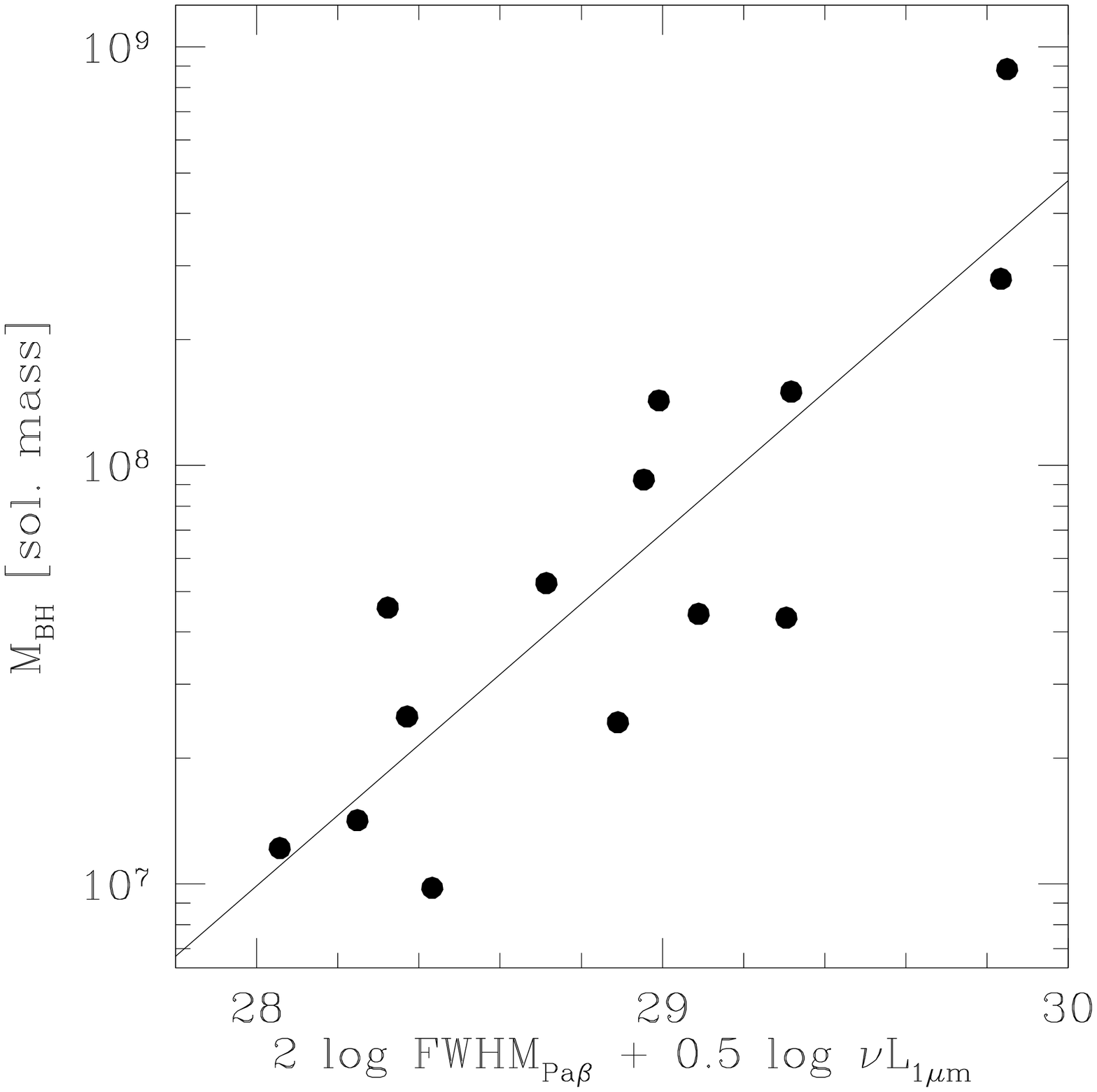}
}
\caption{\label{bhmass} Black hole mass determined from reverberation
  campaigns versus the near-IR virial product between the width of the
  Pa$\beta$ broad component and the continuum luminosity at 1
  $\mu$m. Based on the observed correlation (solid line) we have
  estimated the black hole mass of sources without reverberation
  results.}
\end{figure}

The AGN continuum blueward of $\sim 1$ $\mu$m is believed to be
emitted by the accretion disc, which is considered the main source of
ionising flux producing the broad emission lines. In order to test if
this component is indeed seen in our data, we have calculated
accretion disc spectra. We have assumed a steady geometrically thin,
optically thick accretion disc, in which case the emitted flux is
independent of viscosity and each element of the disc face radiates
roughly as a blackbody with a characteristic temperature depending
only on the mass of the black hole, $M_{\rm BH}$, the accretion rate,
$\dot{M}$, and the radius of the innermost stable orbit
\citep[e.g.,][]{Peterson, FKR}. We have adopted the Schwarzschild
geometry (non-rotating black hole) and for this the innermost stable
orbit is at $r_{\rm in} = 6 \cdot r_{\rm g}$, where $r_{\rm g}$ is the
gravitational radius defined as $r_{\rm g} = G M_{\rm BH}/c^2$, with
$G$ the gravitational constant and $c$ the speed of
light. Furthermore, we have assumed that the disc is viewed face-on.

The accretion disc spectrum is fully constrained by the two
quantities, mass and accretion rate of the black hole. Two thirds of
our sample have black hole masses derived from reverberation mapping
campaigns. For the remainder (7 sources) we have estimated this
quantity by applying the virial theorem $M_{\rm BH} \propto v^2 r/G$,
where $v$ and $r$ are the velocity and radial distance of an orbiting
particle, respectively, to the near-IR. Using the width of the
Pa$\beta$~broad component (denoted ${\rm FWHM}_{\rm Pa\beta}$,
published in Paper I) as a measure of $v$ and the square-root of the
continuum luminosity at 1 $\mu$m (denoted $\nu L_{1\mu m}$) as a
surrogate for $r$, Fig. \ref{bhmass} shows that for the sources with
reverberation mapping results the black hole mass correlates with the
near-IR virial product. The observed correlation is $\log M_{\rm BH} =
0.84 \cdot (2{\rm FWHM}_{\rm Pa\beta} + 0.5\nu L_{1\mu m}) -
16.58$. The accretion rate can be obtained directly from an
approximation of the accretion disc spectrum to the data. We show our
results in Fig. \ref{SEDfit}, left panels (dotted lines) and list the
relevant values in Table \ref{fitstab}.

The calculated accretion disc spectrum approximates well the AGN
continuum slope blueward of $\sim 1$ $\mu$m in all high-redshift
($z\ga0.1$) sources, where the host galaxy contribution to the total
flux is negligible at all wavelengths. In particular the near-IR
spectral part ($\sim 0.8 - 1$~$\mu$m, covered by the IRTF spectrum) is
well reproduced, which is free from major contaminating components
such as, e.g., strong \FeII~emission and the `small blue bump'
\citep{Grandi82, Wills85} sometimes found in the optical spectral
part. However, most important for our future near-IR reverberation
programme is the result that the accretion disc emission can still
dominate at the large wavelength of $\sim1$~$\mu$m. This means that we
observe the flux of the ionising component directly in the near-IR,
which will allow us to determine the AGN state without optical
spectroscopy.

In the lower redshift sources, the host galaxy contribution is
expected to be minimal in the near-IR spectrum, since these
observations were taken through a relatively narrow slit, and in the
bluer part of the optical spectrum, since at these frequencies the AGN
emission lies well above that of the host galaxy. Indeed, in both
these frequency regimes the calculated accretion disc spectrum is
always a satisfactory approximation. However, excess host emission is
often observed at the red end of the optical spectrum, which has the
effect of flattening the spectrum (in a logarithmic $\nu f_\nu$ versus
$\nu$ plot). Nevertheless, subtraction of sufficient host galaxy flux
leads also in this spectral range to an alignment of the data with the
expected accretion disc emission.

\subsection{The hot dust}

\begin{figure}
\centerline{
\includegraphics[scale=0.4]{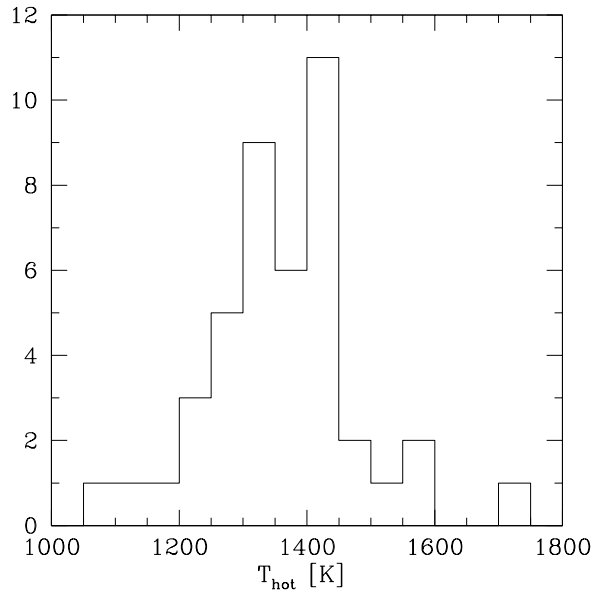}
}
\caption{\label{thot} Histogram of the temperature of the hot dust.}
\end{figure}

\begin{figure}
\centerline{
\includegraphics[scale=0.4]{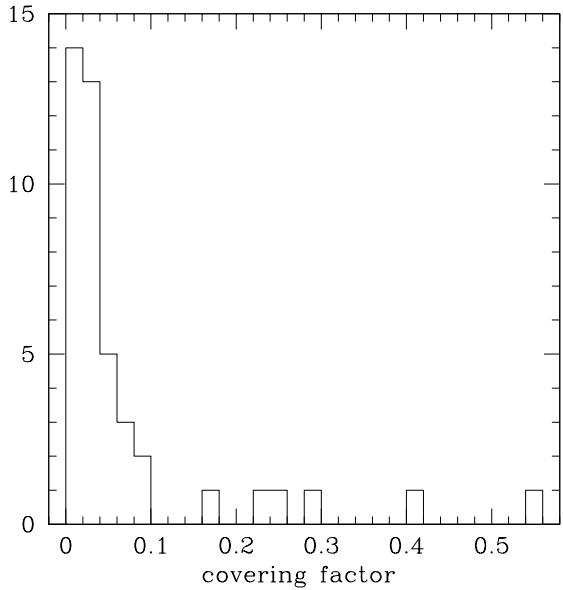}
}
\caption{\label{cov} Histogram of the hot dust covering factor $C
  \approx 0.4 \cdot (\nu L_{\rm hot}/\nu L_{\rm acc})$, where $\nu
  L_{\rm hot}$ and $\nu L_{\rm acc}$ are the peak luminosities of the
  hot blackbody and accretion disc spectrum, respectively.}
\end{figure}

\begin{figure}
\centerline{
\includegraphics[scale=0.4]{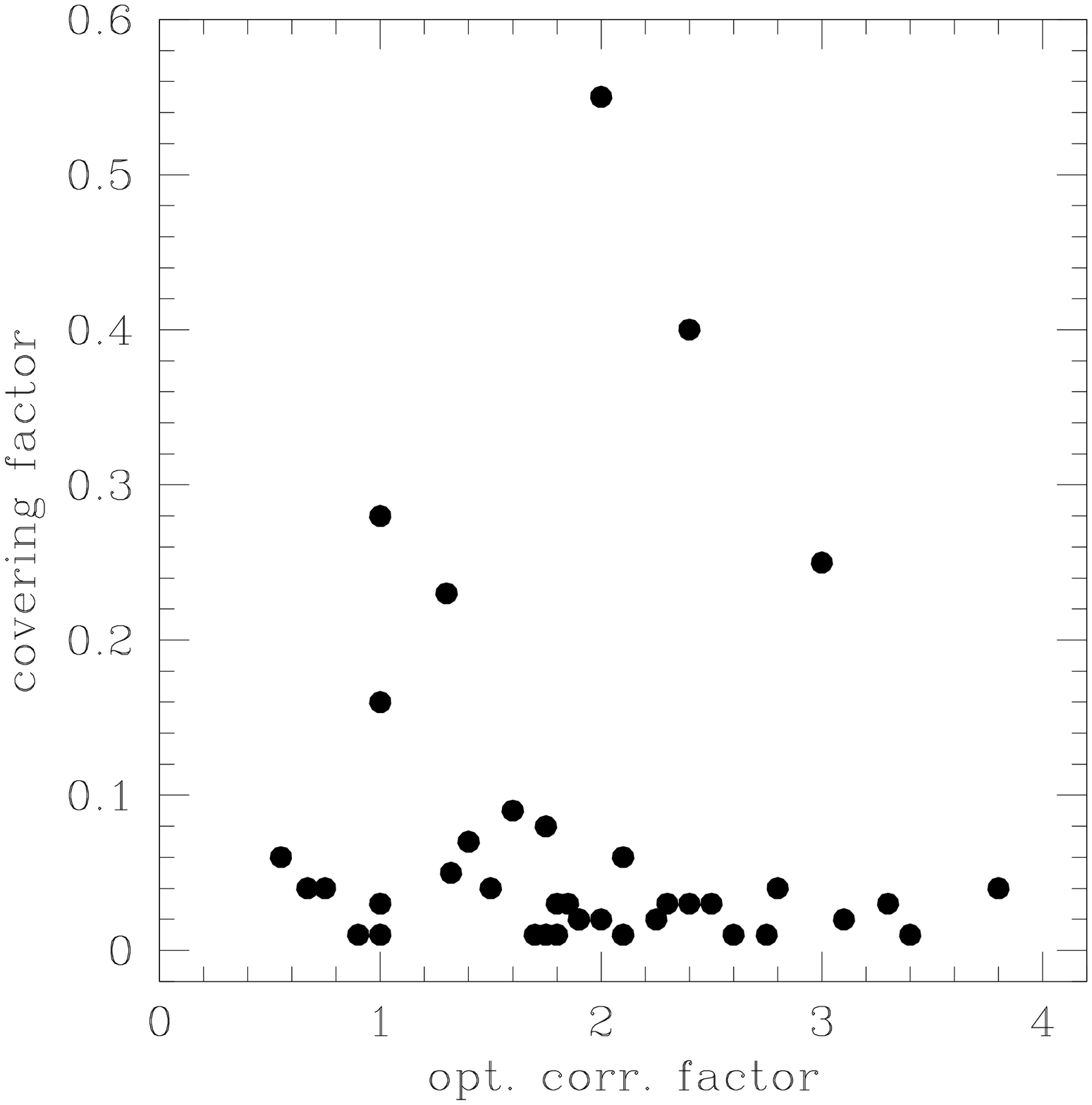}
}
\caption{\label{covcor} The hot dust covering factor versus the
  correction factor applied to the optical spectrum.}
\end{figure}

The AGN continuum redward of $\sim 1$ $\mu$m is believed to be
produced by the hottest part of a dusty torus, which surrounds the
central ionising source and obscures the BELR for lines of sight close
to the plane of the accretion disc. Assuming this component is indeed
seen in our data, we have subtracted from the total spectrum that of
the accretion disc and have fitted to the result a blackbody
spectrum. For this purpose we have used the C routine MPFIT
\citep[version 1.1;][]{mpfit}, which solves the least-squares problem
with the Levenberg-Marquardt technique, and have fitted for the
temperature and flux scaling. We have included in the fit only the
continuum part of the near-IR spectrum, i.e., we have excluded
emission lines, and rebinned it to $\Delta \log \nu = 0.01$ Hz. We
show our results in Fig. \ref{SEDfit}, right panels (dotted lines) and
list the relevant values in Table \ref{fitstab}.

A hot blackbody spectrum appears to approximate well the near-IR AGN
continuum in all our sources. In particular, the typical curvature of
such a spectrum is evident in our spectra due to their relatively
large wavelength coverage. This behaviour was noted also by
\citet{Rod06}, whose near-IR spectrum of one broad-line AGN (Mrk~1239)
covered an even larger wavelength range of $0.8-4.5$ $\mu$m. In this
respect, we note that the blackbody curvature is not evident in the
original spectra (without the accretion disc component subtracted),
which resemble rather a single power-law. The resulting temperatures
for the hot blackbody component are in the range of $T_{\rm hot} \sim
1100 - 1700$ K, which are typical values of the dust sublimation
temperature for most astrophysical grain compositions \citep[$\approx
1000 - 2000$ K;][]{Sal77}. The overall temperature distribution is
relatively narrow and has a well-defined mean of $\langle T_{\rm hot}
\rangle = 1365\pm18$ K (Fig. \ref{thot}). Similarly narrow temperature
distributions were obtained previously by AGN studies using low- and
medium-resolution near-IR spectra \citep{Kob93, Rif09} and more
recently by studies based on high-resolution near-IR photometric and
interferometric observations \citep{Kish07, Kish09, Kish10}.

This well-defined peak of the blackbody temperatures is close to the
sublimation temperature of silicate dust grains \citep{Kimu02}. More
refractory dust types are known, notably carbonaceous dust,
e.g. graphite, which can survive up to $\sim 2000$ K. The absence of
such hot dust, except in one object (NGC~5548), suggests that either
carbonaceous dust is rare in AGN or that some other mechanism than
dust sublimation sets the maximum dust temperature. Dust formation is
critically dependent on the numerical ratio of carbon to oxygen atoms
in the parent gas; most pair up to form CO, and do not form dust
\citep{Whittet}. An imbalance in C/O atom numbers biases dust
formation strongly toward either carbon-rich or oxygen-rich (including
silicates). The $\sim 1400$ K peak in AGN suggests an oxygen-rich
environment from which the dust formed.

The strength of our data set is that it allows us to observe
simultaneously the accretion disc and the hot dust
emission. Therefore, we can derive for the first time meaningful
covering factors for the dusty obscurer in AGN. If the ultraviolet
radiation from the accretion disc emitted into the solid angle,
$\Omega$, defined by the dust distribution is completely absorbed and
re-emitted in the infrared, the dust covering factor is $C =
\Omega/4\pi=\int_{\rm hot} L_\nu d\nu/\int_{\rm acc} L_\nu d\nu
\approx 0.4 \cdot (\nu L_{\rm hot}/\nu L_{\rm acc})$, where $\nu
L_{\rm hot}$ and $\nu L_{\rm acc}$ are the peak luminosities of the
hot blackbody and accretion disc spectrum, respectively
\citep[e.g.,][]{Barv87, Gra94}. Note that whereas $\nu L_{\rm hot}$
lies only slightly outside the observed spectral range and, therefore,
is well-constrained by the data, the accretion disc peak luminosity is
strongly model-dependent. In Fig. \ref{cov} we show the distribution
of hot dust covering factors. We obtain values in the range of $C \sim
0.01 - 0.6$ and a mean of $\langle C \rangle = 0.07\pm0.02$. Our
average value is a factor of $\sim 6$ lower than the average {\it
  total} dust covering factor obtained by \citet{San89} ($\langle C
\rangle = 0.40\pm0.01$), who considered the ratio between the entire
integrated infrared luminosity (in the frequency range $\nu \sim
10^{12} - 10^{14.5}$ Hz) and the integrated accretion disc luminosity
for a large ($\sim 100$ sources) sample of bright quasars.

The relatively low hot dust covering factors could be a result
introduced by the correction factors that we applied to the optical
spectra (Section \ref{agnsed}), i.e., we have artificially raised the
accretion disc flux relative to the hot dust emission. In order to
address this concern, we have plotted in Fig. \ref{covcor} the
covering factors versus the optical correction factors. Whereas we see
an envelope in the sense that the higher the optical correction
factor, the lower the highest covering factor, overall we observe
large hot dust covering factors ($C\ga0.1$) in only a few cases
(mostly in Ark~120 and NGC~4151) or, put differently, AGN states
cluster around relatively low $C$ values independent of the optical
correction factor.

\subsection{Additional continuum components}

The observed AGN SEDs can be explained mainly by the sum of an
accretion disc spectrum and emission from hot dust. However, two
additional continuum components are expected to be present, which
albeit weaker could alter the AGN continuum slope. The first is the
diffuse continuum (thermal emission and scattering) from the BELR,
which can be strong due to the relatively high gas densities found in
these regions. The Balmer jump and continuum ($\lambda < 3646$~\AA)
were identified in AGN spectra long ago \citep[e.g.,][]{Mal82} and
predicted in early photoionization models as well
\citep[e.g.,][]{Wills85}. If the BELR is emitting a Balmer continuum,
it is also emitting a Paschen continuum ($\lambda < 8205$~\AA), the
strength of which could be important throughout the optical and
near-IR spectrum. Other, weaker, recombination continua (Brackett,
etc.) will also be present within the IR. \citet{Kor01} used the
photoionization code \cloudy~\citep{Cloudy} and calculated the full
diffuse BELR continuum assuming clouds with simple distributions in
gas density and incident ionizing flux. They found that after
correcting for host galaxy light $\sim 20\%$ of the flux at
5100~\AA~may be due to this component, which is dominated by the
thermal continua (nearly 40\% at the Balmer jump; see Korista \& Goad
for details). In a representation of the SED as in Fig. \ref{SEDfit}
($\log \nu f_\nu$ vs. $\log \nu$) the relative contribution of this
component is expected to strengthen with increasing wavelength as the
accretion disc spectrum falls, resulting in a flattening of the
near-UV to near-IR AGN continuum spectrum as compared to an accretion
disc alone.

A second additional component is due to the finite albedo of the same
dusty clouds emitting the thermal IR continuum. \citet{Kor98}
investigated the effective albedo of a variety of gas clouds
potentially present within the BELR. These contributions are included
in the diffuse continuum just described. They also presented the
effective albedo of a high column density, dusty cloud with parameters
grossly approximating the clouds producing most of the hot dust
emission. They found a $\sim 20\%$ albedo spanning the wavelength
range of $\lambda = 2500$~\AA~--~1 $\mu$m that declines significantly
for wavelengths outside. While the detailed wavelength-dependent
albedo depends on the grain size distributions and composition, their
overall amplitudes should not differ greatly. Although the
spatially-integrated scattered nuclear light will depend on factors
such as the geometry of the scattering region and the observer's
viewing angle, contributions of $\sim 10\%$ to the AGN continuum are
possible. Due to the finite grain albedo, the dusty torus will
contribute additional light to the near-UV to near-IR spectrum that
otherwise would not be expected from its thermal emission alone. As
with the contribution from the diffuse continuum light from the BELR,
this contribution will act to flatten (in $\log \nu f_\nu$ vs. $\log
\nu$) the AGN continuum.

Due to the uncertainties in the absolute flux calibration and the
relative normalizations of our near-IR and optical spectra, we have
not attempted to isolate the contributions of the diffuse BELR
continuum and the dust scattering of nuclear light to the AGN
continuum. However, doing so should provide additional physical and
geometrical constraints on the BELR and inner dusty torus. Here we
only point out that the strongest contributor to the diffuse BELR
continuum, namely the Balmer jump and continuum are clearly visible in
most of our optical spectra, forming together with \FeII~the so-called
`small blue bump' \citep{Grandi82, Wills85}.

\section{The AGN continuum variability}

\begin{figure}
\centerline{
\includegraphics[scale=0.4]{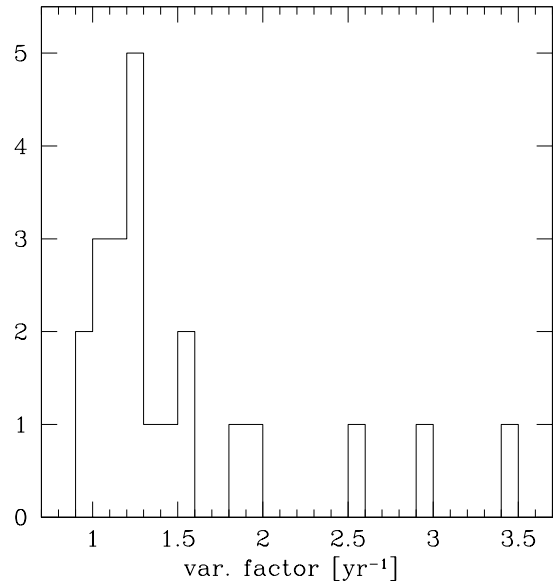}
}
\caption{\label{var} Histogram of the variability factor measured at 1
  $\mu$m and calculated over a fixed period of one year.}
\end{figure}

\begin{figure}
\centerline{
\includegraphics[scale=0.4]{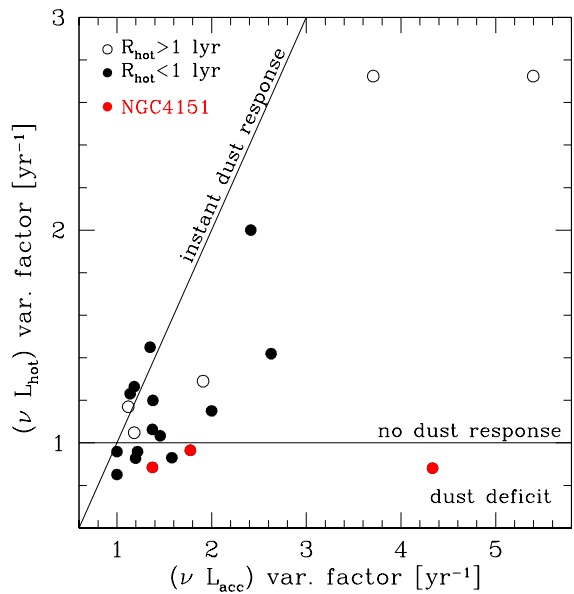}
}
\caption{\label{dustrev} Change in hot dust luminosity versus change
  in accretion disc luminosity over a fixed period of one year. The
  horizontal solid line indicates the locus of equal hot dust
  luminosity in both AGN states, with a deficit of hot dust observed
  in the high state for sources below this line. The diagonal solid
  line marks the locus of equality. Open and filled circles indicate
  sources with average hot dust radii greater and lower than one light
  year, respectively.}
\end{figure}

Both AGN continuum components sampled by our spectra are expected to
be variable. The main cause of variability will be a change in the
accretion rate, which will lead to a change in the dust
illumination. However, although the variability behaviour of the
accretion disc has been studied extensively by optical reverberation
campaigns, only few sources have been observed in dust reverberation
programmes so far \citep[e.g.,][]{Glass92, Nel96, Okn99, Glass04,
  Min04, Sug06, Kosh09}. Therefore, the exact location of the dusty
obscurer as well as its geometry remain uncertain. In this section we
discuss the first results of our coordinated optical and near-IR
reverberation campaign. Based on these results, we will select the
final sample for monitoring and develop a suitable observational
strategy.

We have multiple observation epochs available for 15/21 sources listed
in Table \ref{fitstab}, of which five sources have more than two
epochs. Defining the variability factor as the flux change at 1 $\mu$m
relative to the lowest-flux state in a fixed period of one year, we
obtain the distribution shown in Fig. \ref{var}. Note that this
variability factor is different from that listed in Table
\ref{fitstab} (column (6)), which refers to the total period between
two epochs, and was calculated assuming uniform variability between
the lowest-flux epoch and the epoch with a higher flux. The most
variable of our sources are 3C~273, HE~1228$+$013 and NGC~4151, which
have at least one epoch with a one-year variability factor of $\ga2$,
followed by PG~0844$+$349, Ark~120, NGC~5548 and NGC~4593, with
variability factors of $\sim 1.5-2$. The source Mrk~817 shows no
variability and our three least variable sources are IRAS~1750$+$508,
Mrk~509 and H~2106$-$099.

With our future near-IR reverberation programme we will be able to
study also the variability behaviour of the hot dust and its
dependence on the state of the ionising source. A first look gives
Fig. \ref{dustrev}, where we have plotted the hot dust variability
factor versus the accretion disc variability factor, with both values
calculated for a fixed period of one year. Assuming that the hot dust
absorbs and re-emits all the accretion disc luminosity, we have then
estimated the average radius of the hot dust component as $R_{\rm hot}
= \sqrt{\nu L_{\rm acc}/4 \pi \sigma T_{\rm hot}^4}$, where $\sigma$
is the Stefan-Boltzmann constant, and have separated sources into
those having $R_{\rm hot}>1$ and $<1$ light year (open and filled
circles, respectively).

Two important results become evident from Fig. \ref{dustrev}. Firstly,
for most sources we observe a response of the hot dust to the change
in accretion disc luminosity, which is always lower than the latter,
as expected for a response lag. Most interestingly, however, is that a
dust response is observed even in those sources with estimated hot
dust radii well above one light year, i.e., for which we would expect
no response. The only exception in this group is the source
IRAS~1750$+$508. Secondly, for a few sources we observe a deficit of
hot dust in the high state. This effect is most pronounced in the
source NGC~4151, for which it was observed also by \citet{Kosh09}
using long-term optical and near-IR imaging and attributed to dust
destruction.

\section{Summary and conclusions}

We have used four epochs of quasi-simultaneous (within two months)
near-IR and optical spectroscopy of 23 broad-line AGN to study the
continuum spectral shape around 1 $\mu$m. Our main results can be
summarized as follows.

\vspace*{0.2cm}

(i) The accretion disc spectrum appears to dominate the flux at $\sim
1$~$\mu$m, which allows us to derive a new relation that can be used
to estimate AGN black hole masses. It is based on the near-IR virial
product, defined here as the product between the width of the
Pa$\beta$ broad emission line and the integrated 1~$\mu$m continuum
luminosity. The dominance of the accretion disc spectrum at such long
wavelengths means that the AGN state can be determined directly from
the near-IR spectrum, making simultaneous optical spectroscopy for a
reverberation programme unnecessary.

(ii) An adequate subtraction of (in particular optical) host galaxy
light reveals that the AGN continuum in the rest-frame frequency range
of $\nu \sim 10^{14} - 10^{15}$ Hz can be approximated by the sum of
mainly two emission components, a hot dust blackbody and an accretion
disc spectrum.

(iii) For the hot dust component we derive temperatures in the range of
$T_{\rm hot} \sim 1100 - 1700$~K, which are typical values of the dust
sublimation temperature, with a mean of $\langle T_{\rm hot} \rangle =
1365\pm18$ K. This mean value is close to the sublimation temperature
of silicate dust grains, indicating that either carbonaceous dust is
rare in AGN or that some other mechanism than dust sublimation sets
the maximum dust temperature. The resulting hot dust covering factors
are relatively low and in the range of $C \sim 0.01 - 0.6$, with a
mean of $\langle C \rangle = 0.07\pm0.02$.

(iv) Our preliminary variability studies have identified promising
candidates for a future near-IR reverberation programme. Our three
most variable sources in the near-IR are 3C~273, HE~1228$+$013 and
NGC~4151. Furthermore, we have studied the response of the hot dust
emission to changes in the accretion disc flux. Most sources show the
expected time lag, but a few sources have a deficit of hot dust in the
high state, which indicates dust destruction.

\vspace*{0.2cm}

In our future work we will study the variability of the near-IR broad
emission lines and constrain their physical conditions using detailed
photoionisation models. In the longer term we plan to image our sample
with current and future near-IR interferometers \citep{Elvis02}.

\section*{Acknowledgments}

We are indebted to Brad Peterson for stimulating discussions and his
help with the general project. We thank Perry Berlind and Mike Calkins
for the FAST observations and Susan Tokarz and Nathalie Martimbeau for
the reduction of these data. M. C. B. acknowledges financial support
by the National Science Foundation (grant AST-0604066). This research
has made use of the NASA/IPAC Extragalactic Database (NED), which is
operated by the Jet Propulsion Laboratory, California Institute of
Technology, under contract with the National Aeronautics Space
Administration.

\bibliography{references}

\appendix

\section{Original and Corrected Spectral Energy Distributions}


\begin{figure*}
\centerline{
\includegraphics[scale=1.0]{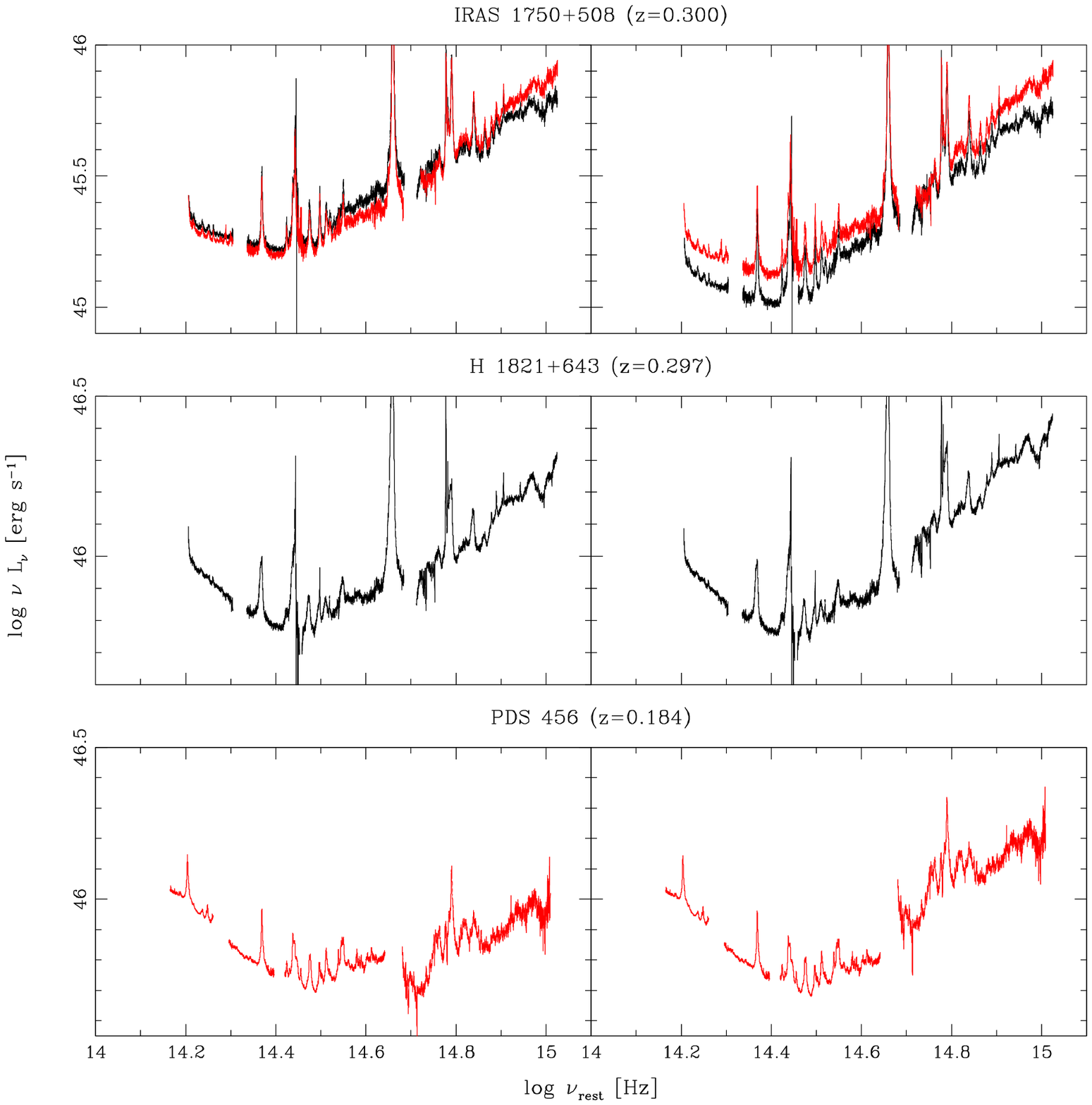}
}
\caption{\label{SEDsub} Rest-frame spectral energy distributions for
  the IRTF observing runs in 2004 May (black), 2006 January (green),
  2006 June (red), and 2007 January (blue). Left and right panels show
  the original data and the data after host galaxy subtraction and
  spectral alignment were applied, respectively. See text for more
  details.}
\end{figure*}

\setcounter{figure}{1}
\begin{figure*}
\centerline{
\includegraphics[scale=1.0]{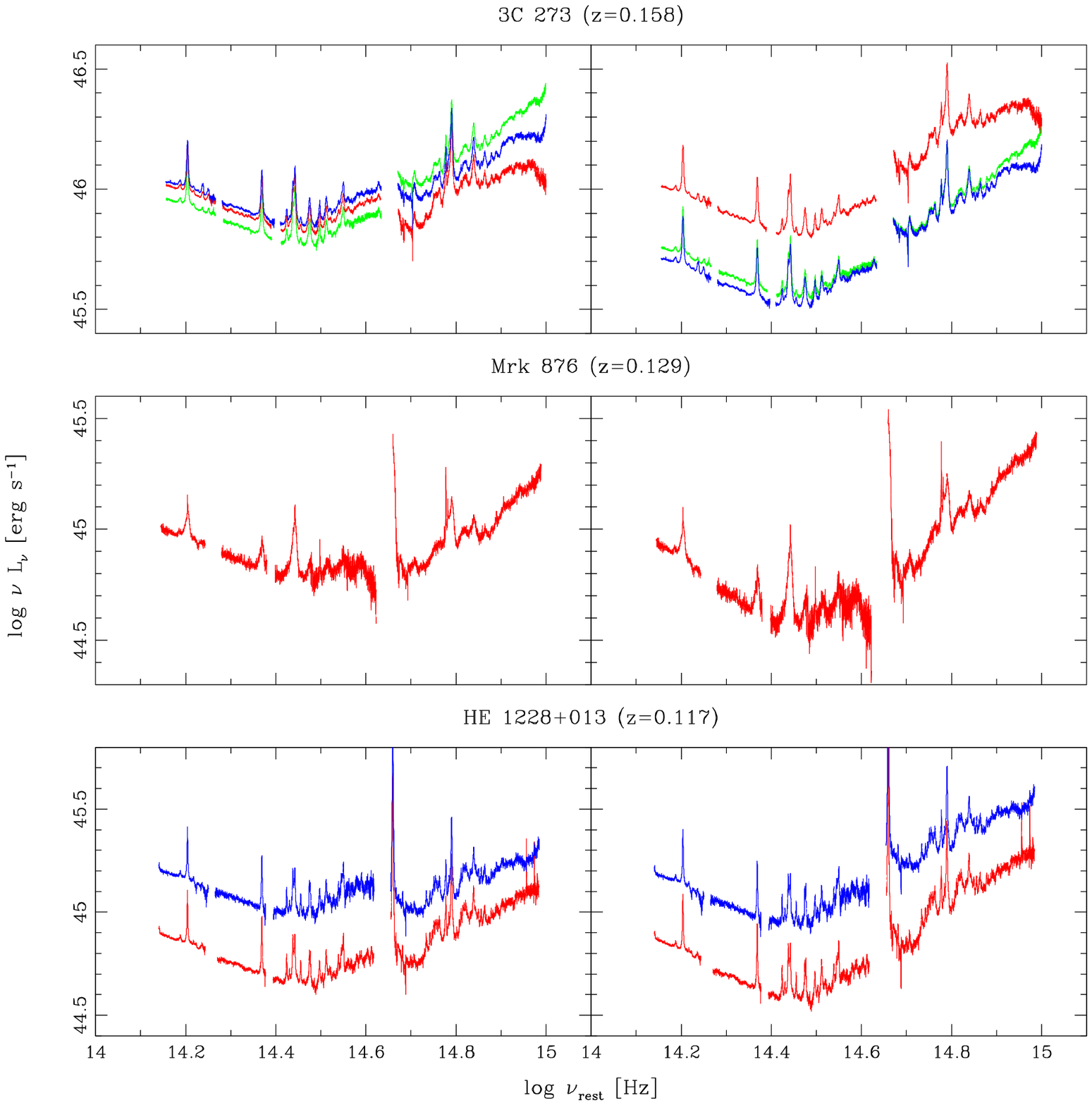}
}
\contcaption{}
\end{figure*}

\setcounter{figure}{1}
\begin{figure*}
\centerline{
\includegraphics[scale=1.0]{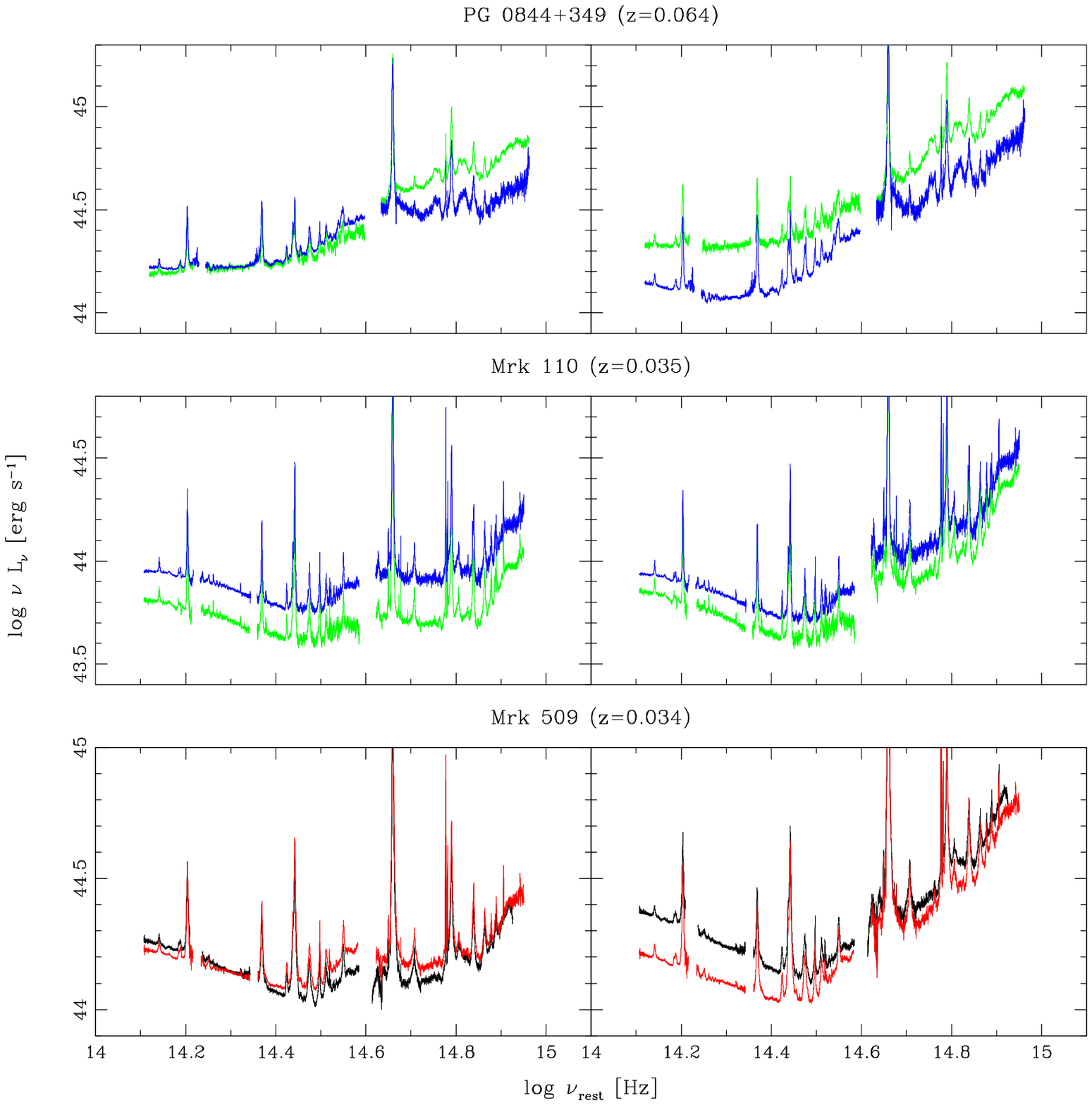}
}
\contcaption{}
\end{figure*}

\setcounter{figure}{1}
\begin{figure*}
\centerline{
\includegraphics[scale=1.0]{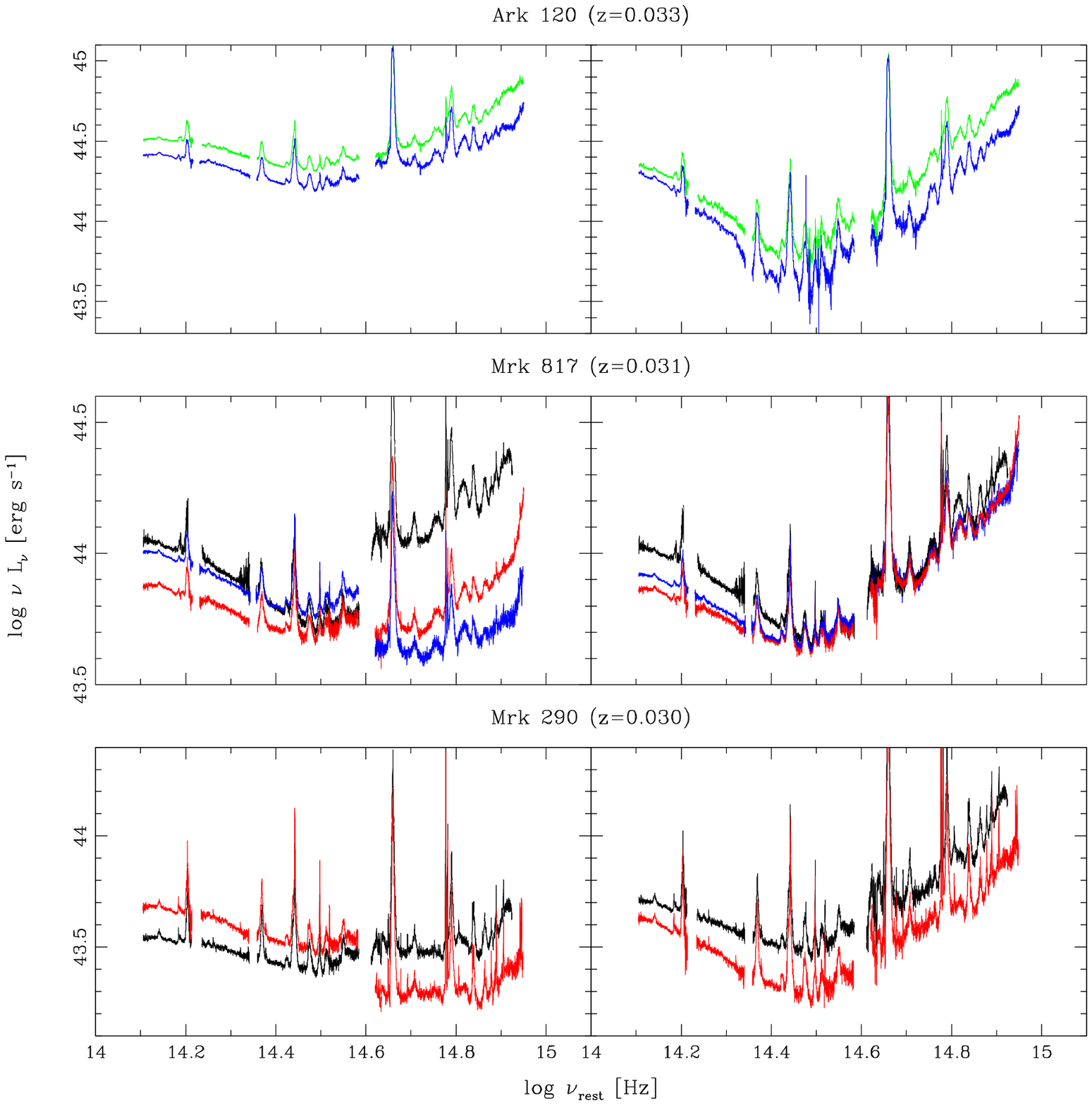}
}
\contcaption{}
\end{figure*}

\setcounter{figure}{1}
\begin{figure*}
\centerline{
\includegraphics[scale=1.0]{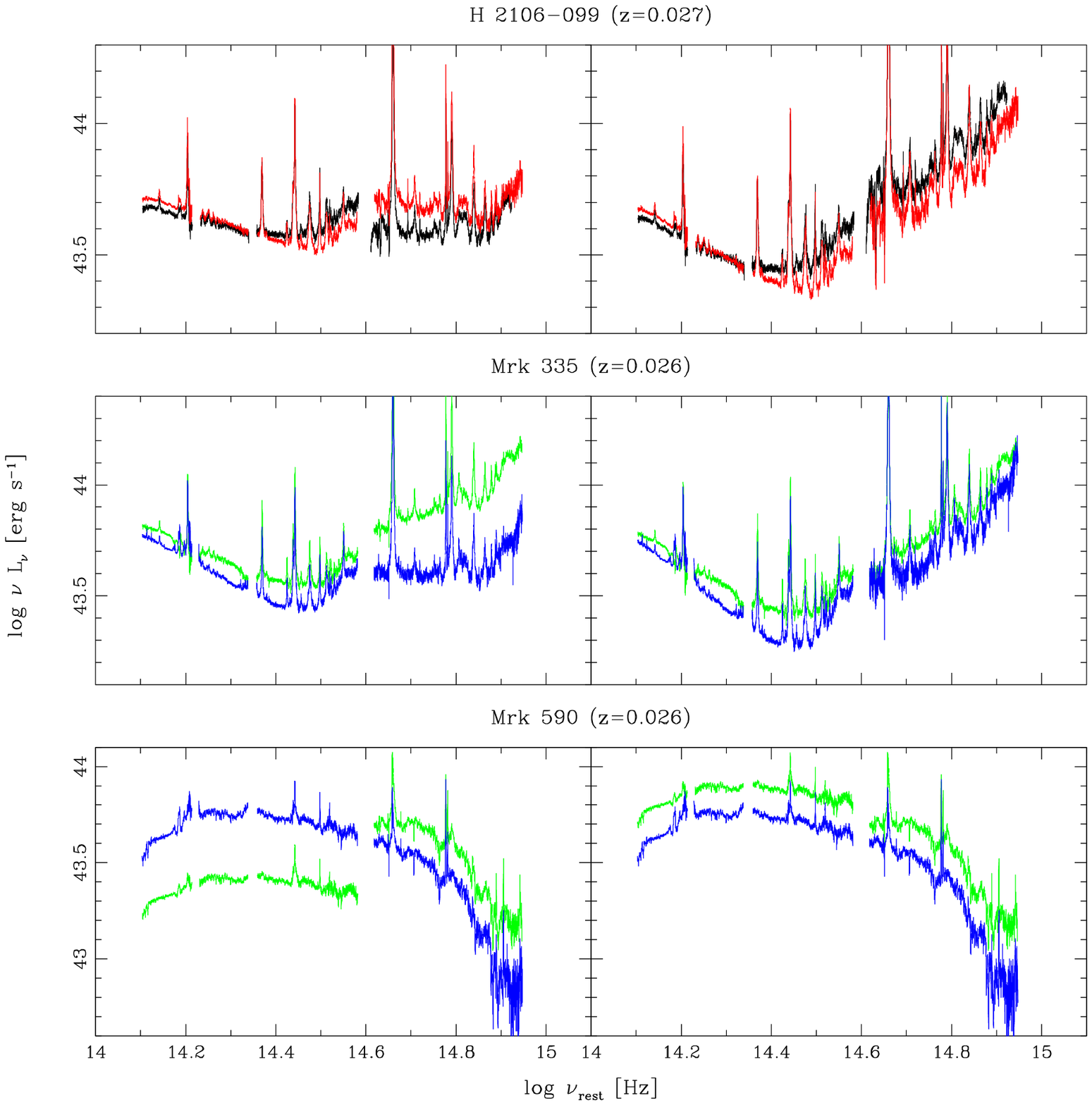}
}
\contcaption{}
\end{figure*}

\setcounter{figure}{1}
\begin{figure*}
\centerline{
\includegraphics[scale=1.0]{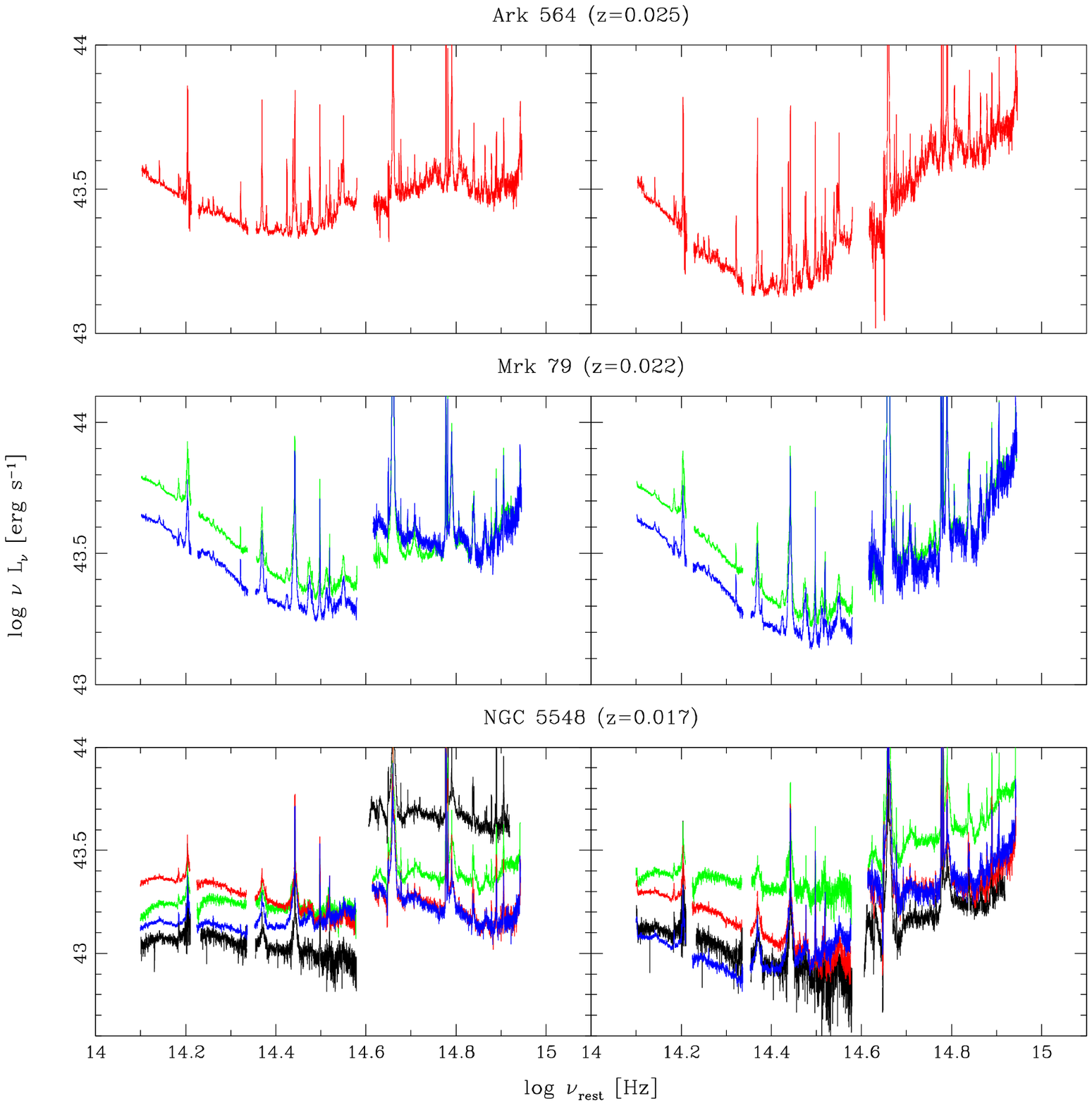}
}
\contcaption{}
\end{figure*}

\setcounter{figure}{1}
\begin{figure*}
\centerline{
\includegraphics[scale=1.0]{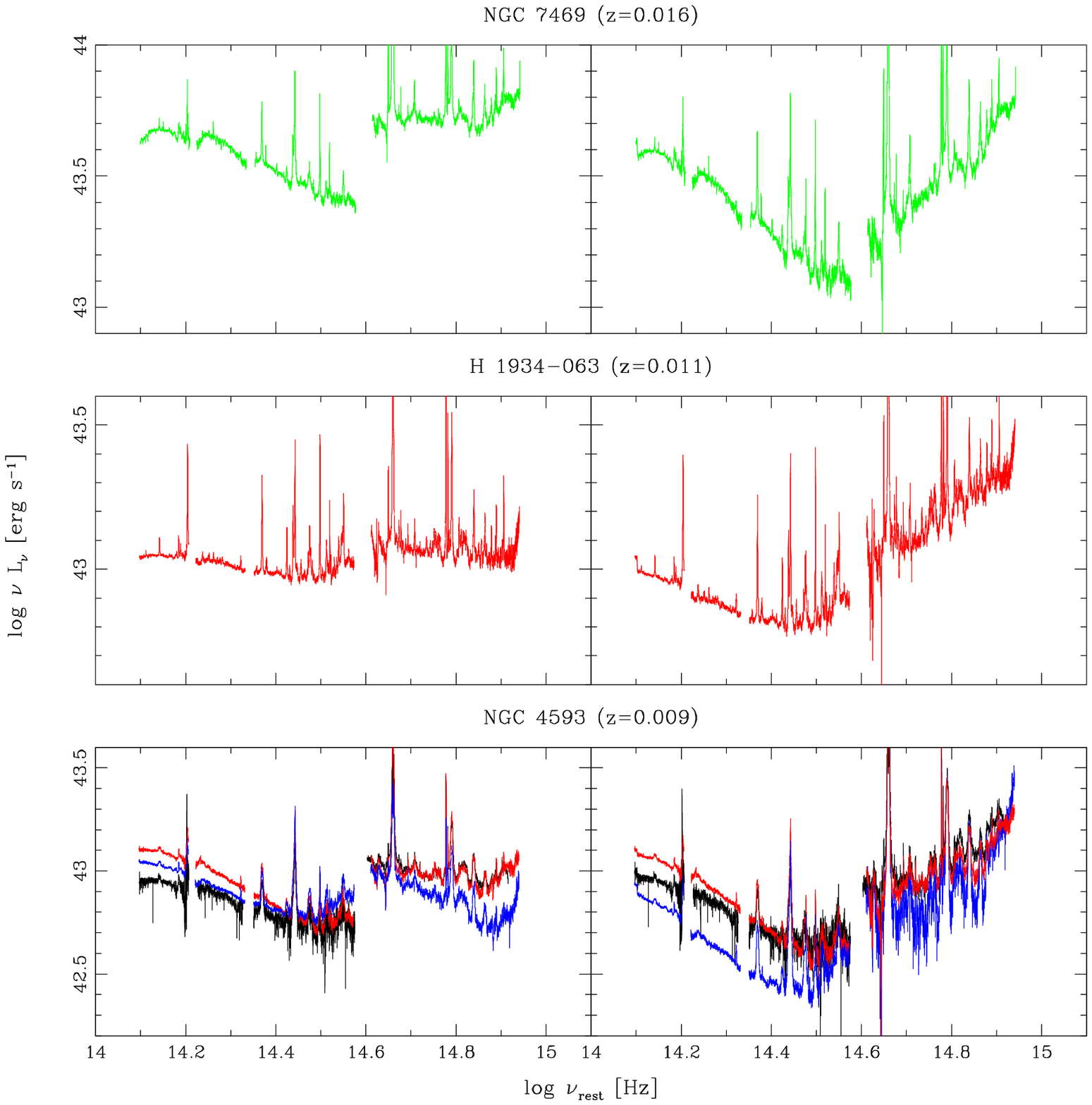}
}
\contcaption{}
\end{figure*}

\setcounter{figure}{1}
\begin{figure*}
\centerline{
\includegraphics[clip=true, bb=17 334 590 715, scale=1.0]{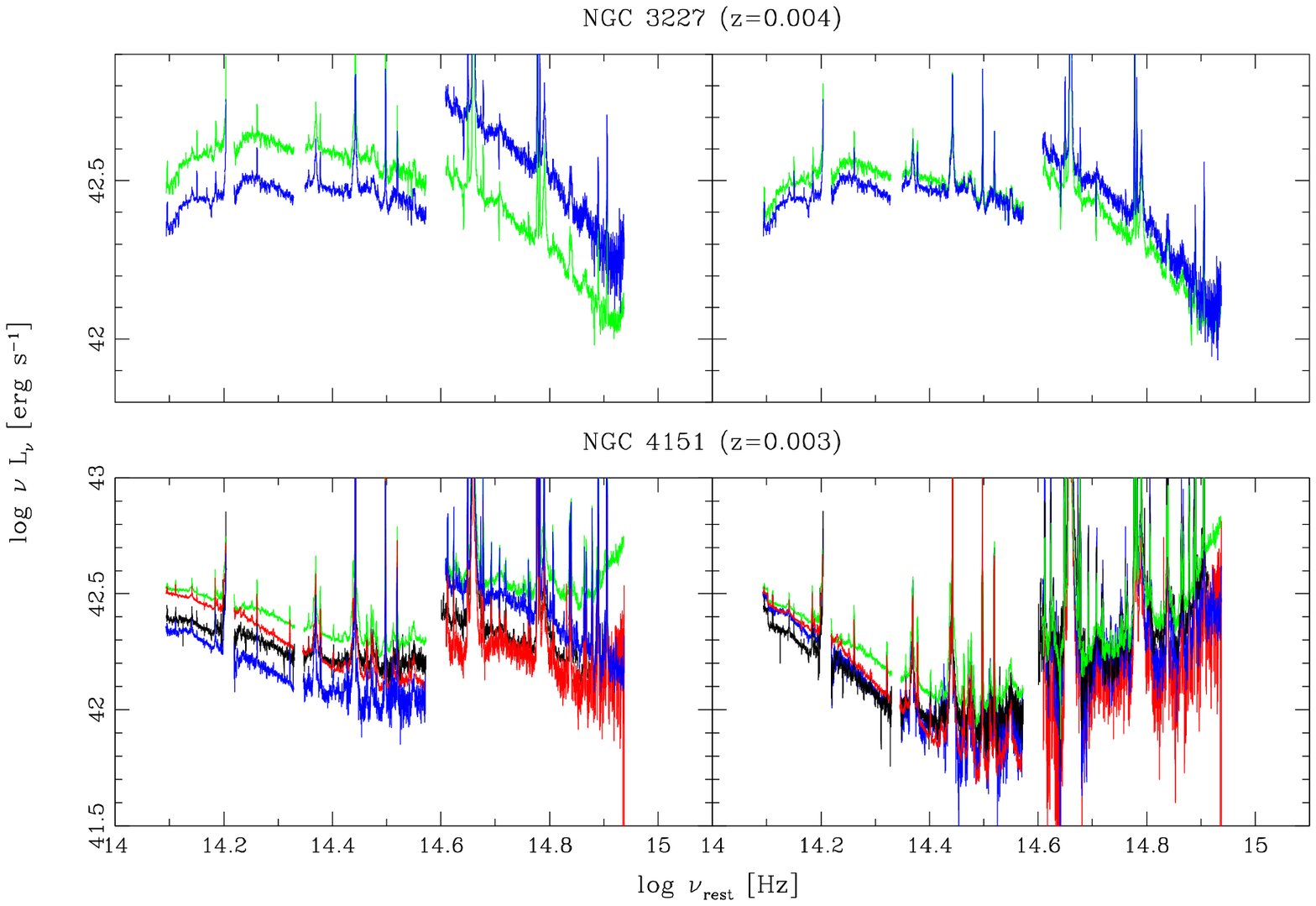}
}
\contcaption{}
\end{figure*}

\bsp
\label{lastpage}

\end{document}